\documentclass[manuscript,screen,acmsmall]{acmart}

\usepackage{amsfonts,amssymb}
\usepackage{booktabs}
\newcommand{\ignore}[1]{}
\usepackage{xspace}

\newcommand{\ie}{\emph{i.e.,}\xspace}

\newcommand{\eg}{\emph{e.g.,}\xspace}

\newcommand{\bx}{\mathbf{x}}
\newcommand{\by}{\mathbf{y}}
\newcommand{\bz}{\mathbf{z}}

\newcommand{\bF}{\mathbf{F}}
\newcommand{\bW}{\mathbf{W}}

\newcommand{\bQ}{\mathbf{Q}}

\newcommand{\bK}{\mathbf{K}}
\newcommand{\bV}{\mathbf{V}}
\newcommand{\bR}{\mathbf{R}}

\usepackage{multirow}
\usepackage{multicol}
\usepackage{caption}
\usepackage{subcaption}
\usepackage{latexsym}
\usepackage{mflogo}
\usepackage{chngpage}

\usepackage{array}
\usepackage{amsthm}
\usepackage[ruled,linesnumbered]{algorithm2e}

\AtBeginDocument{%
  \providecommand\BibTeX{{%
    \normalfont B\kern-0.5em{\scshape i\kern-0.25em b}\kern-0.8em\TeX}}}

\setcopyright{acmcopyright}
\copyrightyear{2018}
\acmYear{2018}
\acmDOI{10.1145/1122445.1122456}

\acmConference[Woodstock '18]{Woodstock '18: ACM Symposium on Neural
  Gaze Detection}{June 03--05, 2018}{Woodstock, NY}
\acmBooktitle{Woodstock '18: ACM Symposium on Neural Gaze Detection,
  June 03--05, 2018, Woodstock, NY}
\acmPrice{15.00}
\acmISBN{978-1-4503-XXXX-X/18/06}

\begin{document}

\title{Curriculum Pre-Training Heterogeneous Subgraph Transformer for Top-$N$ Recommendation}

\author{Hui Wang$^\dagger$}\thanks{$^\dagger$Equal Contribution}
\affiliation{%
  \institution{School of Information, Renmin University of China}}
\email{hui.wang@ruc.edu.cn}

\author{Kun Zhou$^\dagger$}
\affiliation{%
  \institution{School of Information, Renmin University of China}}
\email{francis_kun_zhou@163.com}

\author{Wayne Xin Zhao*}
\affiliation{%
  \institution{Gaoling School of Artificial Intelligence, Renmin University of China. Beijing Key Laboratory of Big Data Management and Analysis Methods}}
\email{batmanfly@gmail.com}
\thanks{$^*$Corresponding author.}

\author{Jingyuan Wang}
\affiliation{
  \institution{School of Computer Science and Engineering, Beihang University}}
\email{bianshuqing@buaa.edu.cn}

\author{Ji-Rong Wen}
\affiliation{
  \institution{Gaoling School of Artificial Intelligence, Renmin University of China. Beijing Key Laboratory of Big Data Management and Analysis Methods}}
\email{jrwen@ruc.edu.cn}

\ignore{
\author{Ben Trovato}
\authornote{Both authors contributed equally to this research.}
\email{trovato@corporation.com}
\orcid{1234-5678-9012}
\author{G.K.M. Tobin}
\authornotemark[1]
\email{webmaster@marysville-ohio.com}
\affiliation{%
  \institution{Institute for Clarity in Documentation}
  \streetaddress{P.O. Box 1212}
  \city{Dublin}
  \state{Ohio}
  \country{USA}
  \postcode{43017-6221}
}

\author{Lars Th{\o}rv{\"a}ld}
\affiliation{%
  \institution{The Th{\o}rv{\"a}ld Group}
  \streetaddress{1 Th{\o}rv{\"a}ld Circle}
  \city{Hekla}
  \country{Iceland}}
\email{larst@affiliation.org}

\author{Valerie B\'eranger}
\affiliation{%
  \institution{Inria Paris-Rocquencourt}
  \city{Rocquencourt}
  \country{France}
}

\author{Aparna Patel}
\affiliation{%
 \institution{Rajiv Gandhi University}
 \streetaddress{Rono-Hills}
 \city{Doimukh}
 \state{Arunachal Pradesh}
 \country{India}}

\author{Huifen Chan}
\affiliation{%
  \institution{Tsinghua University}
  \streetaddress{30 Shuangqing Rd}
  \city{Haidian Qu}
  \state{Beijing Shi}
  \country{China}}

\author{Charles Palmer}
\affiliation{%
  \institution{Palmer Research Laboratories}
  \streetaddress{8600 Datapoint Drive}
  \city{San Antonio}
  \state{Texas}
  \country{USA}
  \postcode{78229}}
\email{cpalmer@prl.com}

\author{John Smith}
\affiliation{%
  \institution{The Th{\o}rv{\"a}ld Group}
  \streetaddress{1 Th{\o}rv{\"a}ld Circle}
  \city{Hekla}
  \country{Iceland}}
\email{jsmith@affiliation.org}

\author{Julius P. Kumquat}
\affiliation{%
  \institution{The Kumquat Consortium}
  \city{New York}
  \country{USA}}
\email{jpkumquat@consortium.net}}

\renewcommand{\shortauthors}{Hui and Zhou et al.}

\begin{abstract}
Due to the flexibility in modelling data heterogeneity, heterogeneous information network (HIN) has been adopted to characterize complex and heterogeneous auxiliary data in top-$N$ recommender systems, called \emph{HIN-based recommendation}. 
HIN characterizes complex, heterogeneous data relations, containing a variety of information that may not be related to the recommendation task. Therefore, it is challenging to effectively leverage useful information from HINs for improving the recommendation performance.
To address the above issue, we propose a Curriculum pre-training based HEterogeneous Subgraph Transformer (called \emph{CHEST}) with new \emph{data characterization}, \emph{representation model} and \emph{learning algorithm}.

Specifically, we consider extracting useful information from HIN to compose the interaction-specific heterogeneous subgraph, containing both sufficient and relevant context information for recommendation.
Then we capture the rich semantics (\eg graph structure and path semantics) within the subgraph via a heterogeneous subgraph Transformer, where we encode the subgraph with multi-slot sequence representations.
Besides, we design a curriculum pre-training strategy to provide an elementary-to-advanced learning process, by which we smoothly transfer basic semantics in HIN for modeling user-item interaction relation.

Extensive experiments conducted on three real-world datasets demonstrate the superiority of our proposed method over a number of competitive baselines, especially when only limited training data is available.

\end{abstract}



\maketitle
\section{Introduction}

\ignore{
\begin{figure}
\includegraphics[width=.9\linewidth]{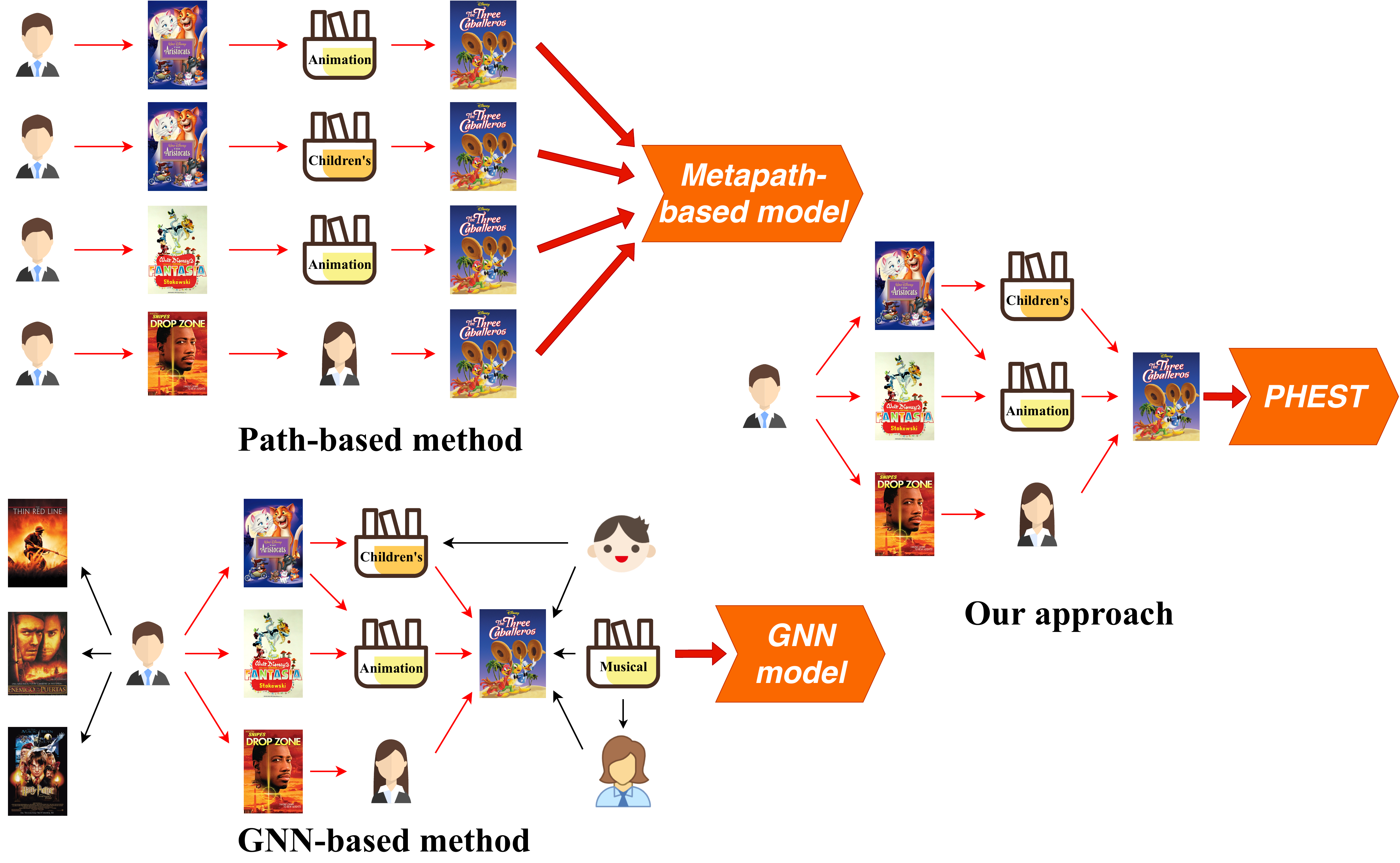}
\centering
\caption{The comparison between our model and existing methods. Our approach constructs the subgraph by collecting edges (marked in red) from sampled paths.}
\label{fig-intro}
\end{figure}}

Online consumptions (\eg purchasing goods and watching movies) have become increasingly popular with the rapid development of Internet services. Meanwhile, people repeatedly encounter the choice problem because of Information Overload~\cite{isinkaye2015recommendation}.
To address such problems, recommender systems (RS) have become an important tool in online platforms, which model users’ preferences on items based on their past interactions.
Due to the complexity of user behavior, recent works utilize various kinds of auxiliary data to improve recommender systems, such as item attributes and user profiles.
These auxiliary data can be considered as important context to understand user-item interaction. It is essential to effectively utilize such context data to improve the recommendation performance~\cite{DBLP:conf/kdd/HuSZY18,DBLP:conf/icdm/Rendle10}.
As a promising approach,  Heterogeneous Information Network (HIN)~\cite{DBLP:journals/tkde/ShiLZSY17,DBLP:journals/pvldb/SunHYYW11,DBLP:conf/wsdm/YuRSGSKNH14}, consisting of multiple types of nodes and edges, has been widely applied to model rich context information in recommender systems. The recommendation task framed in the HIN setting is usually referred to as \emph{HIN-based recommendation}~\cite{DBLP:conf/kdd/HuSZY18,DBLP:conf/kdd/JinQFD00ZS20}.

For HIN-based recommendation task, the most essential problem is how to effectively utilize the rich information in HIN for recommendation task.
A variety of approaches have been proposed to solve this problem, which roughly fall into path-based methods and graph representation learning methods.
Since users and items are connected by paths in HIN, path-based approaches~\cite{DBLP:conf/kdd/HuSZY18,MTRec} mainly focus on sampling paths from HIN and modeling path-level semantics to characterize the user-item interaction relation.
As a widely-used schema, meta-path~\cite{DBLP:journals/pvldb/SunHYYW11} has been used to extract features for depicting the user-item association~\cite{DBLP:conf/wsdm/YuRSGSKNH14,DBLP:conf/kdd/HuFS19}.
By modeling path-based features via similarity factorization~\cite{DBLP:conf/wsdm/YuRSGSKNH14}, co-attention model~\cite{DBLP:conf/kdd/HuFS19} and meta-learning~\cite{DBLP:conf/kdd/Lu0S20}, it is able to improve the recommendation performance~\cite{DBLP:journals/pvldb/SunHYYW11,DBLP:conf/wsdm/YuRSGSKNH14}.
In the other hand, graph representation learning methods~\cite{DBLP:conf/kdd/ZhangSHSC19,DBLP:conf/iclr/KipfW17} consider aggregating features from neighbor nodes in the HIN~\cite{DBLP:conf/www/WangJSWYCY19}, and leverage the graph structure information to learn the data representations~\cite{DBLP:conf/iclr/HuLGZLPL20}.
These methods focus on learning the structural information (\eg edges) in the graph without considering the downstream tasks.
The learned user and item representations will be utilized for predicting the user-item association.

Although existing methods have shown effective to some extent, there are two major challenges that have not been well addressed in HIN-based recommendation. 
First, HIN characterizes complex, heterogeneous data relations, hence it is difficult to extract sufficient contextual semantics and meanwhile avoid incorporating task-irrelevant information from HIN.
Existing approaches either select limited context information from specially designed strategies (\eg path-based methods)~\cite{DBLP:conf/kdd/HuSZY18, MTRec}, or consider the global view that may incorporate noisy information from task-irrelevant nodes and edges (\eg graph representation learning methods)~\cite{DBLP:conf/www/HuDWS20}. There is a need to consider both \emph{relevance} and \emph{sufficiency} in leveraging HIN information for recommendation. Second, HIN is in essence a general data characterization way, 
and it is difficult to design suitable learning strategies to derive task-specific data representations for HIN. 
Existing methods either fully rely on the downstream recommendation task (easy to overfit on training data)~\cite{DBLP:conf/cikm/HuSZY18, DBLP:conf/kdd/HuSZY18}, or employ task-insensitive pre-training strategy (unaware of the final task goal)~\cite{DBLP:journals/corr/abs-2007-03184}. There is a need for a more principled learning algorithm that can more effectively control the learning process with the guidance of the task goal.  



To solve the aforementioned issues, we concentrate on  user-item interaction to design a systematic approach for HIN-based recommendation. Firstly, we design a  more suitable data characterization by introducing \emph{interaction-specific heterogeneous subgraph}, with both sufficient and relevant context information for recommendation. 
Then, we further develop a \emph{heterogeneous subgraph Transformer} that captures rich semantics from interaction-specific subgraphs for the recommendation task.  Furthermore, we propose a \emph{curriculum pre-training} strategy consisting of elementary and  advanced courses (\ie pre-training tasks) to gradually learn from both local and global contexts in the subgraph tailored to the recommendation task. 
The above three aspects jointly ensure that our approach can leverage HIN information for recommendation more effectively .

\ignore{we aim to devise a new framework and more effective training strategy to fully utilize useful information in HIN for recommendation task.
We first extract useful contextual information of the user-item pair from HIN to compose the \emph{interaction-specific heterogeneous subgraph}, and then characterize the rich semantics (\eg path semantics and graph structure) via \emph{heterogeneous subgraph Transformer}.
In this way, we not only explicitly filter out irrelevant information by the subgraph, but also preserve useful context information for recommendation task.
Besides, we propose a \emph{curriculum pre-training} strategy consisting of an elementary course to learn 
local contextual information in the subgraph, and an advanced course to learn global contextual information of the user-item interaction.
As a result, this elementary-to-advanced learning process provides a smooth transition from basic semantics in HIN to context of user-item interaction, which is helpful to transfer useful information from HIN to recommendation task.
}

To this end, in this paper, we propose a Curriculum pre-training based HEterogeneous Subgraph Transformer (called \emph{CHEST}) for HIN-based recommendation. 
First, we construct the interaction-specific heterogeneous subgraph consisting of high-quality paths (derived from meta-paths) that connect a user-item pair, which are extracted from HIN but specifically for recommendation task.
Then, we propose a heterogeneous subgraph Transformer to encode the subgraphs with multi-slot sequence representations. 
It consists of a composite embedding layer to map useful contextual information of nodes (\ie node ID, node type, position in sampled paths, and precursors in the subgraph) into dense embedding vectors, and a self-attention layer to aggregate node and subgraph representations.
Finally, we devise the curriculum pre-training algorithm with both elementary and advanced courses.
The elementary course consists of three pre-training tasks related to node, edge and meta-path, focusing on local context information within the subgraph.
The advanced course is a subgraph contrastive learning task, focusing on global context information at subgraph level for user-item interaction. 

To demonstrate the effectiveness of our approach, we conduct extensive experiments on three real-world datasets.
It shows that our model is able to outperform all baseline models, including path-based methods and graph representation learning methods.
In addition, we perform a series of detailed analysis.
We find that our model can alleviate the data sparsity problem to some extent, and the learned representations after curriculum learning can aggregate into obvious clusters.

Our main contributions are summarized as follows. 
\begin{itemize}
\item We construct the interaction-specific heterogeneous subgraph to leverage useful semantics from HIN to capture the correlations between users and items, and propose the heterogeneous subgraph Transformer to learn useful contextual information from the subgraphs for recommendation task.

\item We devise the curriculum pre-training strategy to learn local and global context information within the interaction-specific heterogeneous subgraph, which gradually extracts useful evidence for user-item interaction to improve the recommendation task.

\item Extensive experiments conducted on three real-world datasets demonstrate the effectiveness of our proposed approach against a number of competitive baselines, especially when only limited training data is available.

\end{itemize}

We organize the following content as follows: Section~\ref{sec-related} discusses the related work of HIN-based recommendation, graph pre-training and curriculum learning.
Section~\ref{sec-problem} and Section~\ref{sec-model} introduce the preliminaries and the proposed approach, respectively.
We present the experiments in Section~\ref{sec-experiment}. 
Section~\ref{sec-conclusion} concludes this manuscript.
\section{Related Work}
\label{sec-related}
Our work is closely related to the studies on
HIN-based recommendation, graph pre-training and curriculum learning.

\subsection{HIN-based Recommendation}
In the literature of recommender systems, early works mainly adopt collaborative filtering (CF) methods to utilize historical interactions for recommendation~\cite{DBLP:reference/sp/KorenB15}, where matrix factorization approach~\cite{DBLP:journals/computer/KorenBV09} and factorization machine~\cite{DBLP:conf/icdm/Rendle10} has shown effectiveness and efficiency in many applications.
Since these methods usually suffer from cold-start problem, many works~\cite{DBLP:conf/kdd/ZhouZBZWY20,DBLP:conf/cikm/ZhouWZZWZWW20} attempt to leverage additional information to improve recommendation performance, including social relation~\cite{DBLP:conf/wsdm/MaZLLK11}, item reviewers~\cite{DBLP:conf/recsys/LingLK14} and knowledge graph~\cite{DBLP:conf/www/WangJSWYCY19}.

To effectively utilize the additional information, some of works focus on using heterogeneous information network (HIN)~\cite{DBLP:conf/kdd/LiARS14, DBLP:conf/kdd/HuSZY18,DBLP:journals/tkde/PhamLCZ16, DBLP:journals/tkde/ShiHZY19} in recommender systems.
In this way, objects are of different types and links among objects represent different relations, which naturally characterize complex objects and rich relations.
A mainstream approach is the path-based methods~\cite{DBLP:conf/kdd/HuSZY18,DBLP:conf/wsdm/YuRSGSKNH14}, where the semantic associations between two nodes are reflected by the paths that connect them.
Various methods are proposed to characterize path-level semantics for recommendation~\cite{DBLP:conf/wsdm/YuRSGSKNH14, DBLP:conf/kdd/HuSZY18, DBLP:conf/kdd/Lu0S20}. 
Early works~\cite{DBLP:journals/pvldb/SunHYYW11,DBLP:journals/tkde/ShiKHYW14} propose several path-based similarity measures to evaluate the similarity of objects in heterogeneous information network, which can also apply in recommendation task.
Furthermore, the concept of meta-path is introduced into hybrid recommender systems~\cite{DBLP:conf/recsys/YuRSSKGNH13}. 
Yu et al.~\cite{DBLP:conf/wsdm/YuRSGSKNH14} take advantage of different types of entity relationships in heterogeneous information network and propose a personalized recommendation framework for implicit feedback dataset.
Luo et al.~\cite{DBLP:conf/icdm/LuoPWL14} propose a collaborative filtering based social recommendation method using heterogeneous relations. 
Hu et al.~\cite{DBLP:conf/kdd/HuFS19} leverage the meta-path based contextual information for capturing user-item correlations.

In recent years, graph representation learning~\cite{DBLP:conf/kdd/0009ZGZNQH19,DBLP:conf/www/WangJSWYCY19} has been introduced to model HINs for improving various downstream applications, including the recommendation task.
Typical works adopt graph neural network (GNN) to aggregate the heterogeneous information from adjacent nodes~\cite{DBLP:conf/kdd/0009ZGZNQH19,DBLP:conf/aaai/WangDLWZ19}, and utilize general-purpose objective to learn node or graph representations.
Zhang et al.~\cite{DBLP:conf/kdd/ZhangSHSC19} propose heterogeneous graph neural network to aggregate feature information of sampled neighboring nodes, and leverage graph context loss to train the model.
Wang et al.~\cite{DBLP:conf/www/WangJSWYCY19} utilize graph attention network to aggregate features from meta-path based neighbors in a hierarchical manner, which mainly focuses on semi-supervised classification task.
Wang et al.~\cite{DBLP:conf/cikm/WangTLSWZ20} learn disentangled user/item representations from different aspects in a HIN, which use meta relations to decompose high-order connectivity between node pairs.

Compared with these studies, our approach combines the merits of path-based and graph representation learning methods to learn recommendation-specific data representations.


\subsection{Graph Pre-training} 
Inspired by the success of pre-training methods in computer vision (CV)~\cite{DBLP:journals/ijcv/RussakovskyDSKS15} and natural language processing (NLP)~\cite{DBLP:conf/naacl/DevlinCLT19}, pre-training technique has been recently applied to graph datasets for improving GNNs~\cite{DBLP:conf/kdd/HuDWCS20}.
The purpose of pre-training on graph is to learn parameters of the model for producing general graph representations, which can be further fine-tuned on different downstream tasks.
It has been shown that pre-training methods have the potential to address scarce labeled data~\cite{DBLP:conf/icml/HendrycksLM19} and out-of-distribution prediction~\cite{DBLP:conf/iclr/HuLGZLPL20}.

As an effective unsupervised pre-training strategy, mutual information maximization~\cite{DBLP:conf/iclr/VelickovicFHLBH19,DBLP:conf/iclr/KongdYLDY20} has been utilized to capture the correlations within the graph (\eg nodes, edges and subgraphs). 
Velickovic et al.~\cite{DBLP:conf/iclr/VelickovicFHLBH19} propose graph information maximization method to learn node representations, which is mindful of the global structural properties of the graph.
Ren et al.~\cite{DBLP:journals/corr/abs-1911-08538} explore the use of mutual information maximization for heterogeneous graph representation learning, which focuses on learning high-level representations based on meta-path.
Hu et al.~\cite{DBLP:conf/iclr/HuLGZLPL20} pre-train an expressive GNN at the level of individual nodes as well as entire graphs so that the GNN can learn useful local and global representations simultaneously.

Besides, contrastive learning and graph generation strategies are also utilized to pre-train GNNs.
Qiu et al.~\cite{DBLP:conf/kdd/QiuCDZYDWT20} utilize contrastive learning to capture the universal network topological properties across multiple networks, which empowers graph neural networks to learn the intrinsic and transferable structural representations.
You et al.~\cite{DBLP:journals/corr/abs-2006-04131} develop a framework about contrastive learning with augmentations for GNN, which can produce graph representations of better generalizability, transferrability and robustness.
Hu et al.~\cite{DBLP:conf/kdd/HuDWCS20} introduces a self-supervised attributed graph generation task to pre-train a GNN so that it can capture the structural and semantic properties of the graph.

Generally, most of these methods aim to learn general node representations based on the whole graph.
As a comparison, we propose a curriculum pre-training strategy to learn recommendation-specific representations, which helps extract useful information from HIN to recommendation task.

\subsection{Curriculum Learning}
Inspired by the human learning process, curriculum learning ~\cite{CL_Bengio} is proposed as a learning paradigm that starts from simple patterns and gradually increases to more complex patterns.
Several studies~\cite{CL_CV,CL_NLP} have shown that this training approach results in better generalization and speeds up the convergence.

Most of the works~\cite{CL_CV,CL_NLP} on curriculum learning focus on feeding training instances to the model from easy to hard.
Gong et al.~\cite{CL_CV} and  utilize curriculum learning in image classification task and show the effectiveness. 
In NLP tasks, Guo et al.~\cite{DBLP:conf/aaai/GuoTXQCL20} and Liu et al.~\cite{CL_QA} improve the performance on machine translation and question answer tasks.
Recently, some works~\cite{CL_CV_task_level,CL_NLP_task_level} explore the curriculum learning strategies in task level, and show that a group of well-designed curriculums are helpful to learn complex knowledge.
Liu et al.~\cite{CL_NLP} train the model with sequentially increased degrees of parallelism to train the model from easy to hard, which achieves significant accuracy improvements over previous non-autoregressive neural machine translation methods.
Sarafianos.~\cite{CL_Object} utilize curriculum learning to transmit the acquired knowledge to the target task~\cite{CL_CV_task_level}.

In this paper, we design a curriculum pre-training strategy to gradually learn from both local and global contexts in the subgraph, which helps our model to leverage HIN information for recommendation more effectively.

\ignore{
We review the related work in the following three aspects, namely HIN-based recommendation, graph pre-training and curriculum learning.

\subsection{HIN-based Recommendation}
As an important research topic, heterogeneous information network (HIN)~\cite{DBLP:conf/kdd/LiARS14, DBLP:conf/kdd/HuSZY18,DBLP:journals/tkde/PhamLCZ16, DBLP:journals/tkde/ShiHZY19} can naturally characterize complex objects and rich relations in recommender systems. 
In the literature, many efforts have been devoted to HIN-based recommendation~\cite{DBLP:journals/tkde/PhamLCZ16,DBLP:journals/tkde/ShiHZY19,DBLP:conf/kdd/ZhaoYLSL17}.
A mainstream approach is the path-based methods~\cite{DBLP:conf/kdd/HuSZY18,DBLP:conf/wsdm/YuRSGSKNH14}, where the semantic associations between two nodes are reflected by the paths that connect them.
Various methods are proposed to characterize path-level semantics for recommendation, including path similarity factorization~\cite{DBLP:conf/wsdm/YuRSGSKNH14}, co-attention model~\cite{DBLP:conf/kdd/HuSZY18} and meta learning~\cite{DBLP:conf/kdd/Lu0S20}.
In recent years, graph representation learning~\cite{DBLP:conf/kdd/0009ZGZNQH19,DBLP:conf/www/WangJSWYCY19} has been introduced to model HINs for improving various downstream applications, including the recommendation task.
Typical works adopt HIN-specific graph convolution network or graph attention mechanism to aggregate the heterogeneous information from adjacent nodes~\cite{DBLP:conf/kdd/0009ZGZNQH19,DBLP:conf/aaai/WangDLWZ19}.
Besides, several studies utilize meta-paths to enhance the performance of GNN models.
These methods either explicitly model meta-paths in the information aggregation process~\cite{DBLP:journals/corr/abs-2007-08294} or aggregate meta-path based neighbors for capturing rich semantics underlying heterogeneous graphs~\cite{DBLP:conf/www/WangJSWYCY19}.

Compared with these studies, our approach has adopted a very different data form, \ie interaction-specific heterogeneous subgraphs, in which high-relevance nodes and edges are kept for each user-item pair.
Another major contribution is that we introduce pretraining to enhance the HIN-based data representations for recommendation. The supervision signal of the recommendation task is usually sparse, and we fully utilize the  intrinsic data correlations of HIN for enhancing the model learning.

\subsection{Graph Pre-training}
\ignore{
Early attempts to pre-train graph representations are skip-gram based network embedding models inspired by Word2vec~\cite{DBLP:conf/nips/MikolovSCCD13}, such as LINE~\cite{DBLP:conf/www/TangQWZYM15}, node2vec~\cite{DBLP:conf/kdd/GroverL16}, and metapath2vec~\cite{DBLP:conf/kdd/DongCS17}. 
Most of them follow the neighborhood similarity assumption that nodes closely connected should be considered similar.}

Inspired by the success of pre-training methods in computer vision~\cite{DBLP:journals/ijcv/RussakovskyDSKS15} and natural language processing~\cite{DBLP:conf/naacl/DevlinCLT19}, pre-training technique has been recently applied to graph datasets for improving GNNs~\cite{DBLP:conf/kdd/HuDWCS20}.
It has been shown that pre-training methods have the potential to address scarce labeled data~\cite{DBLP:conf/icml/HendrycksLM19} and out-of-distribution prediction~\cite{DBLP:conf/iclr/HuLGZLPL20}.
Therefore, contrastive learning~\cite{DBLP:conf/kdd/QiuCDZYDWT20,DBLP:journals/corr/abs-2006-04131} and mutual information maximization~\cite{DBLP:conf/iclr/VelickovicFHLBH19,DBLP:conf/iclr/KongdYLDY20} are leveraged to capture the correlations within the graph (\eg nodes, edges and subgraphs). However, most of these methods aim to learn general node representations based on the whole graph, hence it is likely to propagate noisy information from task-irrelevant nodes into data representations.

More recently, several studies leverage additional supervised signals to  improve the representations in HIN~\cite{DBLP:journals/corr/abs-2007-03184,MTRec}.
They focus on improving the modeling of meta-paths or aggregating of neighbours according the target task. 
As a comparison, we construct a interaction-specific heterogeneous subgraph from HIN to cover the most relevant nodes and edges for the user-item pair.
Our data representation is able to effectively prevent noisy information from task-irrelevant nodes. In addition, with the proposed heterogeneous subgraph Transformer, we can can model  both subgraph-level and path-level semantics reflected in the subgraph, which combines the merits of HIN-based methods and GNN-based methods. 


\subsection{Curriculum Learning}
Inspired by the human learning process, curriculum learning ~\cite{CL_Bengio} is proposed as a learning paradigm that starts from simple patterns and gradually increases to more complex patterns. Several studies
have shown the effectiveness of curriculum learning in the field of computer vision~\cite{CL_CV, CL_Object}, as well as a range of NLP tasks~\cite{CL_NLP, CL_QA}. Most of the works on curriculum learning focus on determining the orders of data~\cite{CL_data_level}. Later, some works explore the curriculum learning strategies in task level~\cite{CL_CV_task_level, CL_NLP_task_level}.}
\section{Problem Formulation}
\label{sec-problem}

\begin{table}
	\caption{Notations and explanations}
	\label{tab:notation}
	\centering
	\setlength{\tabcolsep}{0.9mm}
	\centering
	\begin{tabular}{c||p{96mm}}
	\toprule
	Notation & Explanation \\
	\toprule
	\toprule
	$\mathcal{G}$ & heterogeneous information network \\
	$\mathcal{G}_{u,i}$ & a heterogeneous subgraph connecting user-item pair $\langle u, i \rangle$ \\
	$\mathcal{V}$ & the set of nodes \\
	$\mathcal{E}$ & the set of edges \\
	$\mathcal{A}$ & the set of pre-defined entity types \\
	$\mathcal{R}$ & the set of pre-defined edge types \\
	$\mathcal{U}$ & the set of users \\
	$\mathcal{I}$ & the set of items \\
	$\mathcal{I}_u$ & the set of items that $u$ has interacted with before\\
	$\mathcal{S}$ & the set of slots \\
	$\mathcal{P}$ & the set of meta-paths \\
	\toprule
	$u$ & a user \\
	$r$ & a relation \\
	$o$ & an object \\
	$a$ & an attribute \\
	$\rho$ & a meta-path \\
	$p$ & a path instance \\
	$i$ & an item \\
	$i^{'}$ & a random sampled negative item \\
	$v$ & a node \\
	$C_{v_{t}}$ & the surrounding context for $v_t$ in a heterogeneous subgraph \\
	$\texttt{Pr}(\rho |u,i)$ & the preference score of the user $u$ and item $i$ \\
	$\sigma$ & the sigmoid function \\
	\toprule
	$M_{V}$, $M_{A}$, $M_{S}$, $M_{P}$ & the embedding matrices of node ID, node type, slot and precursor \\
	$E_{V}$, $E_{A}$, $E_{S}$, $E_{P}$ & the embedding matrices of node ID, node type, slot and precursor for a heterogeneous subgraph \\
	$E$ & the composite embedding matrix of a heterogeneous subgraph \\
	$W^{O}$, $W^{Q}_{i}$, $W^{K}_{i}$, $W^{V}_{i}$ & learnable parameter matrices in multi-head self-attention layer\\ $W_{1}$, $W_{2}$ & learnable parameter matrices in point-wise feed-forward network \\
	$W_{N}$, $W_{E}$ & learnable parameter matrices for masked node/edge prediction task \\  
	$b_{1}$, $b_{2}$ & learnable parameter vectors \\
	$F^{l}$ & the input of the $l$-th layer \\
	$F_{u}^{L}$, $F_{i}^{L}$ & the representations of user $u$ and item $i$ from the last self-attention layer \\
	$e_{v}$ & the node ID embedding of $v$ \\
	$z_{\mathcal{G}}$ & the subgraph representation \\
	$head_i$ & the output of the $i$-th head of self-attention layer \\
	\toprule
	$d$ & the embedding dimension \\
	$L$ & the number of layers in the Transformer model \\
	$n$ & the number of nodes in the subgraph \\
	$h$ & the number of head in the multi-head self-attention layer\\
	$\tau$ & the hyper-parameter for softmax temperature \\
 	\bottomrule
	\end{tabular}
\end{table}

\ignore{
In this paper, we consider the recommendation task targeting for implicit feedback~\cite{DBLP:conf/uai/RendleFGS09,DBLP:conf/icdm/Rendle10,DBLP:conf/www/HeLZNHC17}. With $n$ users $\mathcal{U} = {u_1, ..., u_n }$ and $m$ items $\mathcal{I} = {i_1,...,i_m}$, we define each entry $r_{u,i}$ in the user implicit feedback matrix $\bR \in \mathbb{R}^{n \times m}$ as follows: $r_{u,i}=1$  when $\langle u,i \rangle$ interaction is observed, and $r_{u,i}=0$ otherwise. Here the value of 1 in the matrix $\mathbf{R}$ indicates the interaction result between a user and an item, \eg whether a user has watched or rated a movie.

We consider the recommendation task in the setting of Heterogeneous Information Network (HIN), which can be defined as follows:

\emph{Definition 1. \textbf{Heterogeneous Information Network (HIN)}}. A HIN~\cite{DBLP:conf/wsdm/YuRSGSKNH14,DBLP:journals/tkde/ShiLZSY17,DBLP:journals/tkde/ShiHZY19} is defined as a graph $\mathcal{G} = (\mathcal{V}, \mathcal{E})$, in which $\mathcal{V}$ and $\mathcal{E}$ are the sets of nodes and edges, respectively. 
Each node $v$ and edge $e$ are associated with their type mapping functions $\phi: \mathcal{V}\longrightarrow\mathcal{A}$ and $\varphi: \mathcal{E}\longrightarrow\mathcal{R}$, respectively, where $\mathcal{A}$ and $\mathcal{R}$ denote the sets of pre-defined node and edge types, where $|\mathcal{A}|+|\mathcal{R}|>2$.

In HIN, two objects can be connected via different semantic patterns, which are defined as meta-paths~\cite{DBLP:journals/pvldb/SunHYYW11,DBLP:conf/kdd/ZhaoYLSL17}:

\emph{Definition 2. \textbf{Meta-path}}. A meta-path $\rho$ is defined as a path in the form of $a_{1}$ $\xrightarrow{r_1}$ $a_{2}$ $\xrightarrow{r_2}$ \dots $\xrightarrow{r_l}$ $a_{l+1}$ (abbreviated as $a_{1}a_{2}\dots a_{l+1}$), which describes a composite relation $r_1 \circ r_2 \circ \dots \circ r_l$ between object $a_{1}$ and $a_{l+1}$, where ``$\circ$'' denotes the composition operator on relations. Given a meta-path $\rho$, there exist multiple specific paths under the meta-path, which are called path instances denoted by $p$. 

For example, in Fig.~\ref{approach}, user $u_1$ can be connected to item $i_1$ through the paths $u_1$-$i_2$-$u_2$-$i_1$, $u_1$-$i_2$-$a_1$-$i_1$ and $u_1$-$i_2$-$a_2$-$i_1$, which correspond to meta-paths ``$UIUI$'' or  ``$UIAI$''. These paths reflect potential associations between two nodes in HIN. In our task, we mainly focus on the meta-paths starting with a user node and ending with an item node. 

Given the above preliminaries, we are ready to define our task.

\emph{Definition 3. \textbf{HIN-based Recommendation}}. 
Given a HIN $\mathcal{G}$, for each user $u\in\mathcal{U}$, we aim to recommend a ranked list of items that are of interest to $u$ based on her or his historical record $\mathcal{I}_u$, where $\mathcal{I}_u \subset \mathcal{I}$ denotes the set of items that $u$ has interacted with before.

}

A heterogeneous information network (HIN) is a special kind of information network, which either contains multiple types of objects or
multiple types of links.
We consider the recommendation task in the setting of Heterogeneous Information Network.

\emph{Definition 1. \textbf{Heterogeneous Information Network (HIN)}}. A HIN~\cite{DBLP:journals/tkde/ShiLZSY17, HNE_survey_han, HNE_survey_sun_tang} is defined as a graph $\mathcal{G} = (\mathcal{V}, \mathcal{E})$, in which $\mathcal{V}$ and $\mathcal{E}$ are the sets of nodes and edges, respectively. 
Each node $v$ and edge $e$ are associated with their type mapping functions $\phi: \mathcal{V}\longrightarrow\mathcal{A}$ and $\varphi: \mathcal{E}\longrightarrow\mathcal{R}$, respectively, where $\mathcal{A}$ and $\mathcal{R}$ denote the sets of pre-defined entity and edge types, where $|\mathcal{A}|+|\mathcal{R}|>2$.

Recently, HIN has become a mainstream approach to model various complex interaction systems~\cite{DBLP:journals/tkde/ShiLZSY17,DBLP:journals/tkde/PhamLCZ16}. Specially, it has been adopted in recommender systems for characterizing complex and heterogeneous recommendation settings. Based on the above preliminaries, we define our task as following.

\emph{Definition 2. \textbf{HIN-based Recommendation}}. 
In a recommender system, various kinds of information can be modeled by a HIN $\mathcal{G} = (\mathcal{V}, \mathcal{E})$. On recommendation-oriented HINs, two kinds of entities (\ie users and items) together with the relations between them (\ie rating relation) are our focus.
Let $\mathcal{U}\in \mathcal{E}$ and $\mathcal{I}\in \mathcal{E}$ denote the sets of users and items respectively, for each user $u\in\mathcal{U}$, our task is to recommend a ranked list of items that are of interest to $u$ based on her/his historical record $\mathcal{I}_u$, where $\mathcal{I}_u \subset \mathcal{I}$ denotes the set of items that $u$ has interacted with before.

In HIN, two objects can be connected via different semantic patterns, which are defined as \emph{meta-paths}~\cite{DBLP:journals/pvldb/SunHYYW11}. 

\emph{Definition 3. \textbf{Meta-path}}.  A meta-path is defined as a path in the form of $o_{1}$ $\xrightarrow{r_1}$ $o_{2}$ $\xrightarrow{r_2}$ \dots $\xrightarrow{r_l}$ $o_{l+1}$ (abbreviated as $o_{1}o_{2}\dots o_{l+1}$), which describes a composite relation $r_1 \circ r_2 \circ \dots \circ r_l$ between object $o_{1}$ and $o_{l+1}$, where ``$\circ$'' denotes the composition operator on relations. 

\begin{figure}
	\includegraphics[width=.6\linewidth]{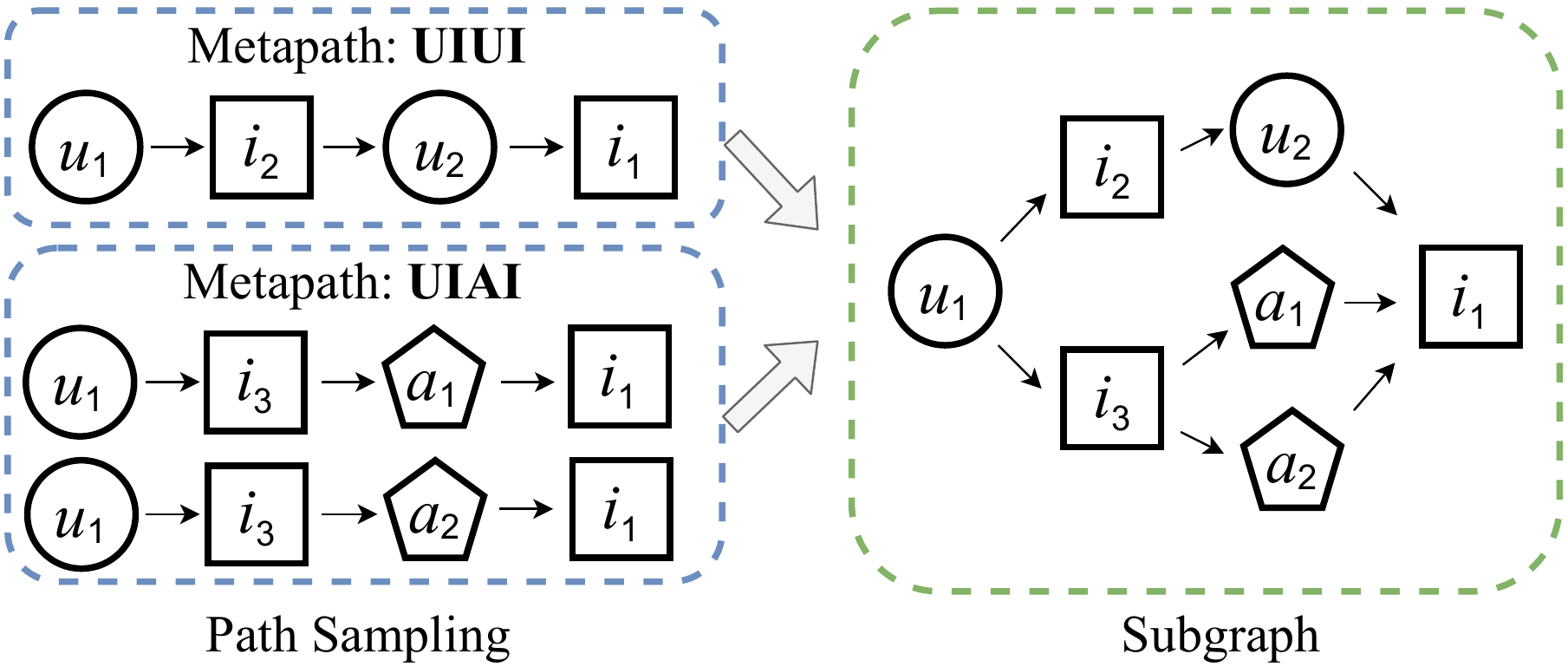}
	\centering
	\caption{The illustration of interaction-specific heterogeneous subgraph, which is constructed by path instances based on meta-paths.}
	\label{fig_sub}
\end{figure}

For a meta-path $\rho$, there exist multiple specific paths following the meta-path, which are called \emph{path instances} denoted by $p$. 
For example, in Figure~\ref{fig_sub}, user $u_1$ can be connected to item $i_1$ through the paths $u_1$-$i_2$-$u_2$-$i_1$, $u_1$-$i_3$-$a_1$-$i_1$ and $u_1$-$i_3$-$a_2$-$i_1$, which correspond to meta-paths ``$UIUI$'' or  ``$UIAI$''. These paths reflect potential associations between two nodes in HIN. 
In our task, we mainly focus on the meta-paths starting with a user node and ending with an item node.

Next, we will present a new curriculum pre-training based heterogeneous subgraph Transformer for this task, which is able to effectively leverage the information reflected in HINs. 
The notations we will use throughout the article are summarized in Table~\ref{tab:notation}.

\ignore{which compose an interaction-specific heterogeneous subgraph for modeling user-item relation.

\emph{Definition 4. \textbf{Interaction-Specific Heterogeneous Subgraph}}.
An interaction-specific heterogeneous subgraph $\mathcal{G}_{u,i}$ is defined as a directed subgraph connecting a user $u$ and an item $i$, which represents necessary context information for modeling user-item interaction.
This subgraph is composed by a set of nodes and edges from the most relevant paths.

To derive relevant and reliable paths between the user and the item nodes, following existing works~\cite{DBLP:conf/kdd/HuFS19,DBLP:journals/pvldb/SunHYYW11}, we pre-define multiple meta-paths to guide the selection of paths. 
For each meta-path, we follow MCRec~\cite{DBLP:conf/kdd/HuSZY18} to adopt a ``\emph{priority}''-based strategy to generate path instances, and keep top-$K$ paths with the highest scores.
The details is given in the appendix.
In Figure~\ref{fig_sub}, we present an example for our interaction-specific heterogeneous subgraph.}
\section{Approach}
\label{sec-model}

\begin{figure*}
	\includegraphics[width=.95\linewidth]{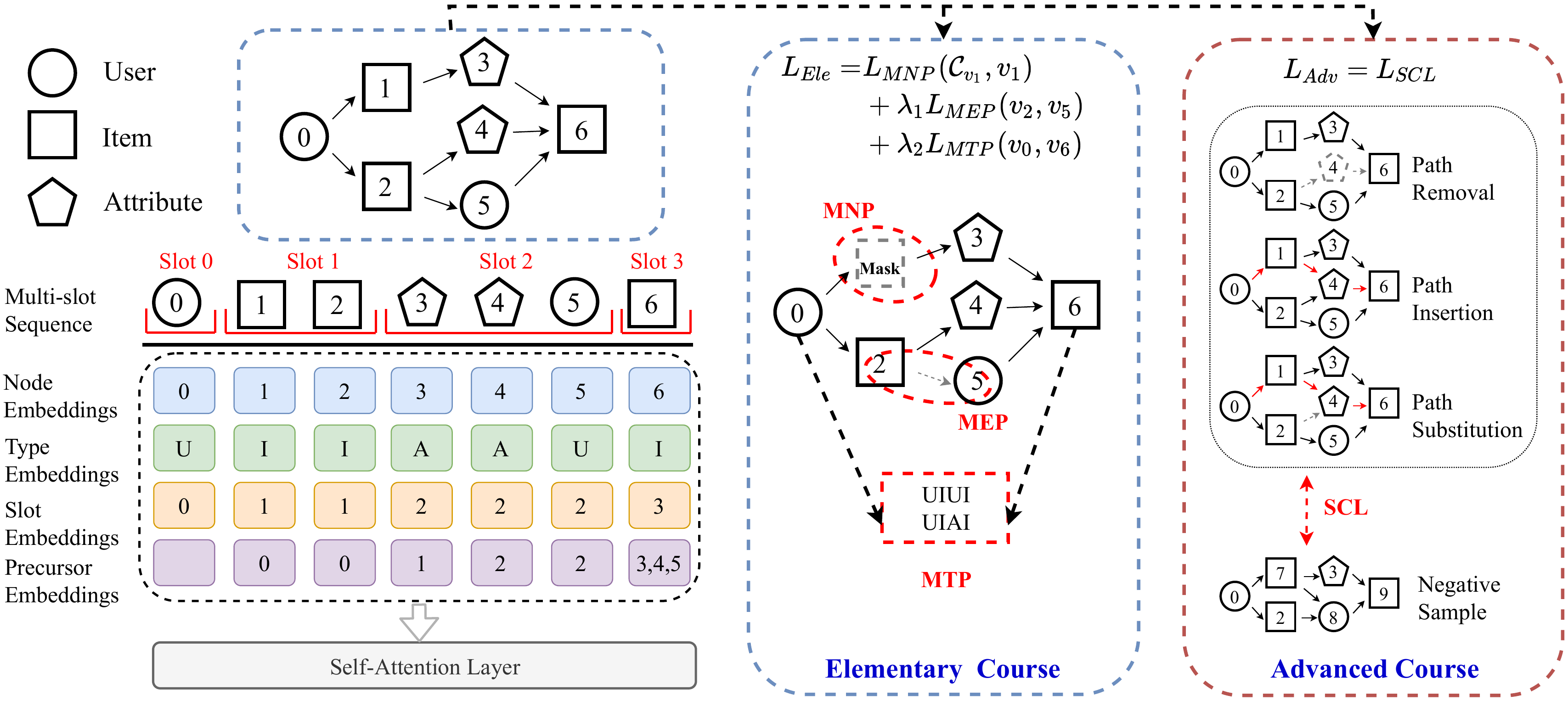}
	\centering
	\caption{The overview of our proposed Transformer model and curriculum pre-training strategy. The elementary courses are three pre-training objectives: (1) Masked Node Prediction (MNP), (2) Masked Edge Prediction (MEP) and (3) Meta-path Type Prediction (MTP). And the advanced course is the Subgraph Contrastive Learning (SCL) task.}
	\label{approach}
\end{figure*}

In this paper, we propose a novel Curriculum pre-training based HEterogeneous subgraph Transformer~(called as \emph{CHEST}) to effectively utilize HIN information for improving the recommendation performance. 
Tailored to the recommendation task, we first construct an interaction-specific heterogeneous subgraph to extract useful contextual information from HIN for the user-item pair, and then design a heterogeneous subgraph Transformer to model this subgraph.
Finally, we introduce curriculum pre-training strategy to learn recommendation-specific representations. Figure~\ref{approach} presents the overall illustration of the proposed CHEST approach. Next, we describe each part in detail.


\subsection{Constructing Interaction-Specific Heterogeneous Subgraph}
In our task, it is essential to leverage useful semantics from HIN to capture the connections between users and items for effective recommendation. 
Different from prior studies~\cite{DBLP:conf/kdd/HuSZY18,DBLP:conf/www/WangJSWYCY19}, we collect the most relevant paths that connect the two nodes. Then, these paths (including nodes and edges) compose a heterogeneous subgraph specially for the user-item pair $\langle u, i \rangle$, denoted by $\mathcal{G}_{u,i}$. We expect such a subgraph
to contain most of the relevant context information for a specific
user-item interaction. 

To derive relevant and reliable paths between two nodes, following existing works~\cite{DBLP:conf/kdd/HuFS19,DBLP:journals/pvldb/SunHYYW11}, we pre-define multiple meta-paths to guide the selection of paths. 
For each meta-path, we follow MCRec~\cite{DBLP:conf/kdd/HuSZY18} to adopt a ``\emph{priority}''-based strategy to generate the path instances. Specifically, we use metapath2vec~\cite{DBLP:conf/kdd/DongCS17} to learn latent vectors of all the nodes, and then a path is evaluated based on the average cosine similarity between the latent vectors of two consecutive nodes on it. Note that the pre-learned latent vectors are only used for path sampling. For each meta-path, we only keep top-$K$ path instances with the highest average similarities. For efficiency consideration, at each step, we walk to nodes in a probabilistic sampling way according to the priority scores of nodes (\ie its similarity with the incoming node). Such an approximate way can reduce the time complexity for constructing heterogeneous subgraphs in practice.

In Figure~\ref{fig_sub}, we present an example for our interaction-specific heterogeneous subgraph for user $u_1$ and item $i_1$, where we consider two types of meta-paths ``$UIUI$'' or  ``$UIAI$''.
For each meta-path, we obtain the corresponding path instances from the HIN by the "priority"-based sampling strategy.
In detail, we acquire the paths $u_1$-$i_2$-$u_2$-$i_1$, $u_1$-$i_3$-$a_1$-$i_1$ and $u_1$-$i_3$-$a_2$-$i_1$ connecting the user-item pair $\langle u_1, i_1 \rangle$ according to meta-paths ``$UIUI$'' and  ``$UIAI$'', respectively.
Finally, we re-connect all the nodes with the edges in these paths, and produce the interaction-specific heterogeneous subgraph as the right part of Figure~\ref{fig_sub}.
With heterogeneous subgraphs, we can explicitly keep the semantics
of multiple meta-paths and model the correlations among nodes across different paths. It is safer and more efficient to aggregate neighboring node information within a compact, relevant subgraph than the entire graph~\cite{DBLP:conf/kdd/HuSZY18,DBLP:journals/corr/abs-2001-05140}, since many of irrelevant nodes in HIN are excluded through the "priority"-based sampling strategy.

\subsection{Heterogeneous Subgraph Transformer}
Given the interaction-specific heterogeneous subgraph for a special user-item pair, we design the heterogeneous subgraph Transformer to capture useful semantics within it, which consists of an embedding layer and a self-attention layer.

\subsubsection{Embedding Layer}
Unlike the embedding mechanism in BERT~\cite{DBLP:conf/naacl/DevlinCLT19} for sequences, we need to effectively model the nodes in the subgraph. To preserve the rich semantics in subgraphs, a key point is how to model the position information (\ie location) of a node and its links with other nodes in the subgraph. For this purpose, we first assign a slot index to a node according to the relative position \emph{w.r.t.} the target user node in a sampled path.
In this way, each node is placed according to its slot index and the original subgraph will be converted into a multi-slot sequence. To  model the links in subgraphs, we  further incorporate a precursor index to trace the precursor in paths for a node.
To facilitate the multi-slot sequence representation, we incorporate four types of node embeddings to preserve the subgraph information:

\ignore{Inspired by the widely-used embedding mechanism in BERT~\cite{DBLP:conf/naacl/DevlinCLT19}, we arrange the nodes of the subgraph in a sequential manner, and devise a series of embedding types to preserve various semantics in the subgraph.
We first assign a slot index to a node according to the relative position in a sampled path. To be specific, the starting user node is assigned to zero, and the ending item node is assigned to the minimum length of multiple involved paths in a subgraph.
Other nodes are ordered according to their position in a path. 
In this way, each node is placed according to its slot index and the original subgraph will be converted into a multi-slot sequence.
To facilitate the multi-slot sequence representation, we incorporate four types of node embeddings to preserve the subgraph information:
}

$\bullet$ \emph{Node ID Embedding}: For each node $v$ in the heterogeneous subgraph $\mathcal{G}_{u,i}$, we maintain an ID embedding matrix $\mathbf{M}_{V} \in \mathbb{R}^{|\mathcal{V}|\times d}$, which projects the high-dimensional one-hot ID representation of a node into low-dimensional dense representation.

$\bullet$ \emph{Node Type Embedding}: 
In HIN, each node is associated with a specific node type $\mathcal{A}$. Therefore, we also maintain a node type embedding matrix $\mathbf{M}_A \in \mathbb{R}^{|\mathcal{A}|\times d}$ to project the one-hot node type representation into dense representation.

$\bullet$ \emph{Slot Embedding}:  
The interaction-specific heterogeneous subgraph is composed by multiple paths. For a node, its distance to the starting node (\ie the target user) is important to consider for the recommendation task. 
Since we have assigned a slot index for each node according to the relative position in the involved paths, we use a slot embedding matrix $\mathbf{M}_S \in \mathbb{R}^{|\mathcal{S}|\times d}$ to project the slot index of nodes into corresponding representations, where $|\mathcal{S}|$ is the number of slots in the subgraph.

$\bullet$ \emph{Precursor Embedding}: 
Although the slot embedding has modeled the relative distance with the starting user node, it cannot capture the adjacent relations between two consecutive nodes in the subgraph. 
Hence, we further add a precursor index to record the preceding nodes of each node in the subgraph.
We maintain a precursor embedding matrix $\mathbf{M}_P \in \mathbb{R}^{n\times d}$ to project the precursor indices of each node into embeddings, where $n$ is the number of nodes in the subgraph.
Since a node may have multiple precursors, we sum the embeddings of the precursor indices as a single vector. 

Based on the above embeddings, we  aggregate them together to produce the node representations in a multi-slot sequence form.
Formally, the node representations of the subgraph is a node embedding matrix $\mathbf{E}\in\mathbb{R}^{|\mathcal{N}|\times d}$, which is composed of four parts:
\begin{align}
	\label{eq-emb}
	\mathbf{E}=\mathbf{E}_{V}+\mathbf{E}_{A}+\mathbf{E}_{S}+\mathbf{E}_{P},
\end{align}
where the four matrices $\mathbf{E}_{V}$, $\mathbf{E}_{A}$, $\mathbf{E}_{S}$ and $\mathbf{E}_{P}$ denote the node ID embedding, node type embedding, slot embedding and precursor embedding, respectively, which are obtained by the look-up operation from $\mathbf{M}_{V}$, $\mathbf{M}_{A}$, $\mathbf{M}_{S}$ and $\mathbf{M}_{P}$, respectively.
It is worth noting that through the above representations, the heterogeneous (\eg node type), path-level (\eg position in the path) and graph-structure (\eg edges in the subgraph) information from subgraph $\mathcal{G}_{u,i}$ have been encoded in the composite embedding matrix $\mathbf{E}$.

\subsubsection{Self-Attention Layer}
Similar to the architecture of Transformer~\cite{DBLP:conf/nips/VaswaniSPUJGKP17}, based on the embedding layer, we develop the subgraph encoder by stacking multiple self-attention layers. A self-attention layer generally consists of two sub-layers, \ie a multi-head self-attention layer and a point-wise feed-forward network. 
Specifically, the multi-head self-attention is defined as:
\begin{align}
	\label{eq-mha}
	\text{MHAttn}(\mathbf{F}^{l})&=[head_{1},head_{2},...,head_{h}]\mathbf{W}^{O}, \\
	head_{i}&=\text{Attention}(\mathbf{F}^{l}\mathbf{W}_{i}^{Q}, \mathbf{F}^{l}\mathbf{W}_{i}^{K}, \mathbf{F}^{l}\mathbf{W}_{i}^{V}),
	\label{eq-head}
\end{align}
where the $\mathbf{F}^{l}$ is the input for the $l$-th layer, when $l=0$, we set $\mathbf{F}^{0}=\mathbf{E}$, and the projection matrix $\mathbf{W}_{i}^{Q}\in\mathbb{R}^{d\times d/h}$, $\mathbf{W}_{i}^{K}\in\mathbb{R}^{d\times d/h}$, $\mathbf{W}_{i}^{V}\in \mathbb{R}^{d\times d/h}$ and $\mathbf{W}^{O}\in\mathbb{R}^{d\times d}$ are the corresponding learnable parameters for each attention head. 
The attention function is implemented by scaled dot-product operation:
\begin{eqnarray}
    \text{Attention}(\bQ,\bK,\bV)=\text{softmax}(\frac{\bQ\bK^{\top}}{\sqrt{d/h}})\bV,
\end{eqnarray}
where $\bQ=\bF^{l}\bW^{Q}_{i}$, $\bK=\bF^{l}\bW^{K}_{i}$, and $\bV=\bF^{l}\bW^{V}_{i}$ are the linear transformations of the input embedding matrix, and $\sqrt{d/h}$ is the scale factor to avoid large values of the inner product.
After the multi-head attention layer, we endow the non-linearity of the self-attention layer by applying a point-wise feed-forward network. 
The computation is defined as:
\begin{align}
	\mathbf{F}^{l} &= [\text{FFN}(\mathbf{F}^{l}_{1})^{\top};\cdots;\text{FFN}(\mathbf{F}^{l}_{n})^{\top}], \label{eq:Fl}\\  
	\text{FFN}(x) &= (\text{ReLU}(x\mathbf{W}_{1}+\mathbf{b}_{1}))\mathbf{W}_{2}+\mathbf{b}_{2}, 
	\label{eq-ffn}
\end{align}
where $\mathbf{W}_{1}$,$\mathbf{b}_{1}$,$\mathbf{W}_{2}$,$\mathbf{b}_{2}$ are trainable parameters. 

Finally, we can compute the representation for the interaction-specific heterogeneous subgraph $\mathcal{G}_{u,i}$ based on the representations at the final self-attention layer as:
\begin{align}
	\bz_{\mathcal{G}_{u, i}} = \text{MLP}(\mathbf{F}^{L}_{u}\oplus \mathbf{F}^{L}_{i}),
	\label{eq-subgraph}
\end{align}
where ``$\oplus$'' denotes the vector concatenation operation, $\mathbf{F}^{L}_{u}$ and $\mathbf{F}^{L}_{i}$ are the representations of user $u$ and item $i$ from the last self-attention layer, which represent the starting user $u$ and ending item $i$ in the subgraph, and $L$ is the number of self-attention blocks.

\subsection{Curriculum Pre-training}
With the above model architecture, we focus on developing an effective representation learning approach that is specially for HIN-based recommendation.
Considering that HIN encodes complex and heterogeneous data relations, our idea is to gradually extract and learn useful information from \emph{local} (\eg node-level) to \emph{global} (\ie subgraph-level) context from interaction-specific heterogeneous subgraphs. Such an idea can be in essence characterized by \emph{curriculum learning}~\cite{CL_Bengio,CL_CV_task_level,CL_NLP_task_level}, which starts from simple tasks or instances and gradually transforms to more complex ones.
Based on this idea, we develop a novel curriculum pre-training strategy that designs both elementary and advanced courses (\ie pre-training tasks) with \emph{increasing difficulty levels}. 

 \ignore{via a curriculum pre-training strategy.
Curriculum learning~\cite{CL_Bengio,CL_CV_task_level,CL_NLP_task_level} is a learning paradigm, which starts from simple/elementary tasks or instances and gradually transfers to more complex/advanced ones.
For HIN-based recommendation task, we consider an elementary course to learn local context information (e.g. node representations) within the interaction-specific heterogeneous subgraph, and an advanced course to learn global context information for the user-item interaction.
This elementary-to-advanced learning process provides a smooth inside-out understanding of the user-item interaction subgraph, which can improve recommendation performance.
}

\subsubsection{Elementary Course}
Elementary course aims to leverage local context information from interaction-specific heterogeneous subgraphs.
We propose to train the proposed heterogeneous subgraph Transformer model with three new tasks, namely masked node prediction, masked edge prediction and meta-path type prediction.
The first two tasks focus on enhancing the node-level representations, while the meta-path type prediction task is designed for capturing path-level semantics for user-item interactions.


$\bullet$ \emph{Masked Node Prediction}:
\label{sec:mnp}
This task is to infer a masked node based on its surrounding context in a heterogeneous subgraph. 
Following the \emph{Cloze} task in BERT~\cite{DBLP:conf/naacl/DevlinCLT19}, 
we randomly mask a proportion of nodes in a heterogeneous subgraph
and then predict the masked nodes based on the remaining contexts.
Assume that we mask node $v_t$ in a multi-slot sequence $\{v_1, \cdots, v_t, \cdots, v_n\}$. We treat the rest sequence $\{v_1, \cdots, \textsc{mask}, \cdots, v_n\}$ as the surrounding context for $v_t$, denoted by $\mathcal{C}_{v_t}$. 
Given the surrounding context $\mathcal{C}_{v_t}$ and the masked node $v_t$, we minimize the Masked Node Prediction (MNP) loss by:
\begin{align}
	L_{MNP}(\mathcal{C}_{v_t}, v_t)=-\log \big(\sigma ( \mathbf{F}_{t}^{\top} \mathbf{W}_{N} \mathbf{e}_{v_{t}} ) - \sigma ( \mathbf{F}_{t}^{\top} \mathbf{W}_{N} \mathbf{e}_{\Tilde{v}} )\big), \label{eq:mnp}
\end{align}
where $\Tilde{v}$ denotes an irrelevant node, $\mathbf{e}_{v_{t}}$ and $\mathbf{e}_{\Tilde{v}}$ denote the node ID embedding for $v_{t}$ and $\Tilde{v}$ respectively,  $\mathbf{W}_{N} \in \mathbb{R}^{d\times d}$ is a parameter matrix to learn and $\mathbf{F}_{t}$ is the learned representation for the $t$-th position using our subgraph encoder as in Eq.~\ref{eq:Fl}.

$\bullet$ \emph{Masked Edge Prediction}:
\label{sec:rep}
The masked edge prediction task is to recover the masked edge of two adjacent nodes based on surrounding context. Similar to masked node prediction, we randomly mask a proportion of edges in the input (\ie removing the precursor index) and then predict the masked edges based on the surrounding contexts.
Formally, the Masked Edge Prediction (MEP) loss for the edge $\langle v_j, v_k \rangle$ can be given as:
\begin{align}
	L_{MEP}(v_j, v_k)=-\log \big(\sigma ( \mathbf{F}_{j}^{\top} \mathbf{W}_{E} \mathbf{F}_{k} ) - \sigma ( \mathbf{F}_{j}^{\top} \mathbf{W}_{E} \mathbf{F}_{k'} )\big), \label{eq:rep}
\end{align}
where $v_{k'}$ is a sampled node that is not adjacent to $v_j$, $\mathbf{W}_{E} \in \mathbb{R}^{d\times d}$ is a parameter matrix to learn, $\mathbf{F}_{j}$, $\mathbf{F}_{k}$ and $\mathbf{F}_{k'}$ are the learned representations for the corresponding positions obtained in the same way as Eq.~\ref{eq:Fl}.

$\bullet$ \emph{Meta-path Type Prediction}:
Since the user-item interaction subgraph is composed by multiple paths,  path-level semantics encode important evidence to explain the underlying reasons why a specific user-item interaction occurs~\cite{DBLP:conf/kdd/HuSZY18,DBLP:conf/cikm/HuSZY18}.
We would like to directly capture the semantics from meta-paths for improving the path semantic information in representations. Specifically, 
we consider meta-path type prediction as a classification task, and introduce the Meta-path Type Prediction (MTP) loss as:
\begin{align}
	\label{eq:mtp}
	L_{MTP}(u,i) = -\sum_{\rho \in \mathcal{P}}\big(y_{u,i,\rho} \cdot \log \text{Pr}(\rho|u,i)+(1-y_{u,i,\rho})\cdot \log (1-\text{Pr}(\rho|u,i))\big),
\end{align}
where $y_{u,i,\rho}$ is a binary label indicating whether there exists a path from the meta-path $\rho$ between $u$ and $i$,
$\mathcal{P}$ is the meta-path set, and $\text{Pr}(\rho|u,i)$ is the probability that the user and item are connected by the meta-path $\rho$, which is defined as:
\begin{align}
	\text{Pr}(\rho|u,i)=\sigma(\text{MLP}(\mathbf{F}^{L}_{u}\oplus \mathbf{F}^{L}_{i})),
\end{align}
where $\text{MLP}(\cdot)$ is a multi-layer perceptron with the sigmoid function as output.

\subsubsection{Advanced Course}
Although the above pre-training tasks have captured local context information (\eg node, edge and path) from the heterogenous subgraph, the global correlations at subgraph level cannot be effectively learned  by elementary course.
To characterize the overall effect of global contexts on recommendation, we devise an advanced course to train the heterogeneous subgraph Transformer with a \emph{Subgraph Contrastive Learning (SCL)} task. 
Based on the original subgraph, the core idea is to augment a number of interaction-specific subgraphs. Then, we apply contrastive learning~\cite{MoCo, SimCLR} to further capture subgraph-level evidence for modeling user-item interaction.
Here, we consider three path-based subgraph augmentation strategies: 

\ignore{
To learn the global context information of the user-item interaction, we devise an advanced course to train our model with a Subgraph Contrastive Learning (SCL) task, which focuses on learning different views of each user-item interaction.
To generate new views of the user-item interaction, we propose three path-level augmentation strategies, in which we remove or insert paths to augment new subgraphs based on the original subgraph. Note that these paths all connect the user-item pair and the inserted path have not been selected to form the interaction-specific subgraph .
}

$\bullet$ \emph{Path Removal}:
It augments new subgraphs by randomly removing a small portion of paths from the original user-item interaction subgraph, which is expected to make the learned representations less sensitive to structural perturbation.

$\bullet$ \emph{Path Insertion}:
It introduces a small proportion of new paths into the original subgraph. 
Many user-item interactions are with sparse connected paths, which is easy to result in over-fitting on the observed data. This task is able to improve the robustness of the recommendation model. 

$\bullet$ \emph{Path Substitution}:
It can be considered as the combination of the path removal and path insertion strategies.
By further enlarging the difference between the augmented subgraphs and original subgraph, this task enforces the model to capture the most fundamental  semantics for user-item interactions. 

Given the target user-item subgraph $\mathcal{G}_{u, i}$ (focusing on user $u$ and $i$), we first augment a set of 
new subgraphs with the above subgraph augmentation strategies, and consider them as \emph{positive subgraphs}, denoted by $\{\mathcal{G}^{+}_{u, i}\}$. While, we consider the subgraphs connecting the same user $u$ with other items $i'$ as \emph{negative subgraphs}, denoted by  $\{\mathcal{G}^{-}_{u, i}\}$.
Following a standard constative learning approach~\cite{SimCLR}, we maximize the difference of augmented positive subgraphs and negative subgraphs, \emph{w.r.t.} the original subgraph: 
\begin{equation}
\begin{aligned}
L_{SCL}(\mathcal{G}, \mathcal{G}^{+}, \{\mathcal{G}^{-}\}) = -\log \frac{\exp \big(\mathrm{sim}(\bz_{\mathcal{G}}, \bz_{\mathcal{G}^{+}})/\tau \big )}{\exp\big({\mathrm{sim}(\bz_{\mathcal{G}}, \bz_{\mathcal{G}^{+}})/\tau}\big)+\sum_{\mathcal{G}^{-}} \exp\big({\mathrm{sim}(\bz_{\mathcal{G}}, \bz_{\mathcal{G}^{-}})/\tau}\big)}, \label{eq:cl}
\end{aligned}
\end{equation}
where $z_{\mathcal{G}}$, $z_{\mathcal{G}^{+}}$ and $z_{\mathcal{G}^{-}}$
are the produced subgraph representations from the heterogeneous subgraph Transformer (Eq.~\ref{eq-subgraph}) for the original subgraph, augmented positive subgraph and augmented negative subgraph (we omit $u$ and $i$ in subscripts for simplicity), respectively,  $\texttt{sim}(\bx,\by)$ denotes the cosine similarity function,  and $\tau$ is a hyper-parameter for softmax temperature.

This constative learning loss enforces the model to learn  subgraph-level semantics for user-item interaction. 
By combining with the elementary course, both local and global context information can be captured in final learned representations. In particular, we schedule the pre-training tasks from two courses in an ``\emph{easy-to-difficult}" order, which is necessary to model complex data relations in HIN.   


\subsection{Learning and Discussion}
In this part, we present the learning and related discussions of our approach for HIN-based recommendation.

\begin{algorithm}[t]
\small
	\caption{The overall training process for the CHEST model.}
	\label{algorithm}
	\LinesNumbered
	\KwIn{
		The heterogeneous information network $\mathcal{G}=(\mathcal{V},\mathcal{E})$, pre-defined meta-paths $\mathcal{P}$, the user set $\mathcal{U}$, the item set $\mathcal{I}$, the user-item historical records $\mathcal{D}={\langle u,i \rangle}$
	}
	\KwOut{The learned node embedding matrix $\mathbf{E}$, the learned parameters of the self-attention layer $\Theta$}
	Use metapath2vec to learn latent vectors of all the nodes in $\mathcal{G}$.\\
	\For{$j = 1 \to |\mathcal{U}|$}{
	    \For{$k = 1 \to |\mathcal{I}|$}{
	        \For{$l = 1 \to |\mathcal{P}|$}{
	            Collect the top-$K$ path instances with the highest average similarities corresponding to the meta-path $\rho_{l}$ that connect the user node $u_j$ and item nodes $i_k$.\\
	        }
	        Merge the collected path instances into a subgraph $\mathcal{G}_{u_j,i_k}$.\\
	    }
	}
	Randomly initialize $\mathbf{E}$ and $\Theta$.\\
	\For{$j = 1 \to |\mathcal{U}|$}{
	    \For{$k = 1 \to |\mathcal{I}|$}{
	        Transform the subgraph $\mathcal{G}_{u_j,i_k}$ for the user-item pair into multi-slot sequence.\\ \label{alg-1}
	        Acquire ID embeddings $\mathbf{E}_{V}$, node type embeddings $\mathbf{E}_{A}$, slot embeddings $\mathbf{E}_{S}$ and precursor embeddings $\mathbf{E}_{P}$ for the nodes in the subgraph $\mathcal{G}_{u_j,i_k}$.\\
		    Acquire the composite embedding matrix $\mathbf{E}$ using Eq.~\ref{eq-emb}.\\
		    Acquire the subgraph representations $\mathbf{F}^{L}$ by multiple self-attention layers using Eq.~\ref{eq-mha}, Eq.~\ref{eq-head}, Eq.~\ref{eq:Fl}, Eq.~\ref{eq-ffn} and Eq.~\ref{eq-subgraph}.\\ \label{alg-2}
		    Pre-train the parameters $\mathbf{E}$ and $\Theta$ using Eq.~\ref{eq:mnp}, Eq.~\ref{eq:rep} and Eq.~\ref{eq:mtp}.\\
		    }
	}
	\For{$t = 1 \to |\mathcal{D}|$}{
	    encode the subgraph $\mathcal{G}_{u_j,i_k}$ using the operations from line~\ref{alg-1} to line~\ref{alg-2}.\\
		Compute $\text{Pr}(u,i)$ using Eq.~\ref{eq:score}.\\
		Fine-tune the parameters $\mathbf{E}$ and $\Theta$ using Eq.~\ref{eq:ft}.\\
	}
	\Return $\mathbf{E}$ and $\Theta$.
\end{algorithm}

\subsubsection{Learning}
The entire procedure of our approach consists of two major stages, namely curriculum pre-training and fine-tuning stages. At the curriculum pre-training stage, we first pre-train our model on the elementary course, consisting of three pre-training objectives to learn 
local context information in the subgraph, then pre-train on the advanced course to learn global context information from HIN.
At the fine-tuning stage, we utilize the pre-trained parameters to initialize the parameters, and then adopt the recommendation task to train our model. Given user $u$ and item $i$, the preference score is calculated by:
\begin{align}
	\label{eq:score}
	\text{Pr}(u,i)=\sigma(\bz_{\mathcal{G}_{u, i}}),
\end{align}
where $\sigma(.)$ is the sigmoid function and $\bz_{\mathcal{G}_{u, i}}$ is the representation for $\mathcal{G}_{u, i}$ defined in Eq.~\ref{eq-subgraph}. We adopt the binary cross-entropy loss as the final objective:
\begin{align}
	L_{rec}(u,i) = -\log \text{Pr}(u,i)-\log(1-\text{Pr}(u, i')),
	\label{eq:ft}
\end{align}
where we pair each ground-truth item $i$ with one (or several) negative item $i'$ that is randomly sampled.
The detailed learning process is shown in Algorithm~\ref{algorithm}.

\subsubsection{Time complexity}
In recommender systems, online service time is more important to consider than offline training time. Once our model has been learned (after pre-training and fine-tuning),  online service time mainly includes the cost of evaluating all the candidate items according to Eq.~\ref{eq:score} and the cost of selecting top items, which is similar to previous neural collaborative filtering methods~\cite{DBLP:conf/www/HeLZNHC17}. A major preprocessing cost lies in the construction of heterogeneous subgraphs for possible user-item pairs. As discussed before, we can pre-compute the priority scores of neighbors for all the nodes. Based on priority scores, we can sample a high-quality path instance in a time roughly as $\mathcal{O}(\bar{L})$ using pre-built efficient data structures such as alias table~\cite{DBLP:conf/kdd/LiARS14} (taking time $\mathcal{O}(1)$ to sample from categorical distributions), where $\bar{L}$ is the average path length. In practice, the number of meta-paths and the number of paths in a subgraph are usually set to small values, so that the number of nodes in a subgraph can be bounded below a reasonable value (\eg 50). 
In this way, our pre-training and fine-tuning costs are similar to train/pre-train Transformer architecture~\cite{DBLP:conf/nips/VaswaniSPUJGKP17,DBLP:conf/naacl/DevlinCLT19} over sequence data, which can be efficient if we use very few self-attention layers or parallelize the computation. 

\subsubsection{Discussion}
Compared with existing work for HIN-based recommendation, our approach has two major differences.
In our approach, data characterization, representation model and learning algorithm are specially designed for user-item interaction based on HIN. 
As for \emph{data characterization}, we introduce interaction-specific heterogeneous subgraph to reduce the incorporation of irrelevant information. Based on such a subgraph structure, we further propose a novel heterogeneous subgraph Transformer as the \emph{representation model}, which can effectively model the subgraph semantics. Furthermore, we propose a novel \emph{learning algorithm} by designing a  curriculum pre-training approach, in which elementary and advanced courses are organized to gradually extract local and global context information from HIN to recommendation task.
The three aspects jointly ensure that our approach can better extract and leverage relevant contextual information from HIN for modeling user-item interaction. 

Our work is related to two categories of models, namely path-based methods~\cite{DBLP:conf/kdd/HuSZY18,DBLP:conf/recsys/YuRSSKGNH13}, graph representation learning methods~\cite{DBLP:conf/kdd/0009ZGZNQH19,DBLP:conf/kdd/QiuCDZYDWT20,DBLP:conf/www/WangJSWYCY19}.
The former category separately models the sampled paths, so that graph-structure or cross-path node correlation can not be explicitly captured. 
Besides, these path-based methods rely on the recommendation task to learn the representations, which may suffer from data sparsity problem and cause overfitting.
As a comparison, our approach construct an interaction-specific heterogeneous subgraph based on high-quality paths, which is able to capture richer semantics from the subgraph structure.
In addition, we propose an elementary-to-advanced curriculum pre-training strategy to gradually learn from both local and global contexts in the subgraph, which is able to learn more effective representations.
Graph representation learning methods aggregate information from neighbouring nodes in the entire HIN and then learn the representations via task-insensitive objectives.
In this way, noisy information from recommendation-irrelevant nodes can be incorporated into the learned representations, and the learning process may be unnecessary for downstream tasks.
In our approach, the interaction-specific heterogeneous subgraph is utilized to incorporate high-quality context information, and meanwhile reduce the the influence of irrelevant nodes and edges.
Then, we design a curriculum pre-training strategy based on the subgraph to learn the user-item association, which is more suitable to the recommendation task.

\ignore{Compared with existing work for HIN-based recommendation task, our approach has two major differences.
	First, our approach proposes an elegant way to balance capturing more information from HIN and avoiding task-irrelevant information.
	Our proposed interaction-specific heterogeneous subgraph is utilized to control the relevance of the contextual information with user-item interaction, and the heterogeneous subgraph Transformer is able to combine the useful features of mainstream methods, consisting of graph-structure and path-level semantics.
	As a comparison, previous HIN-based recommendation methods either neglect some essential information in HIN or inevitably introduce noise information from task-irrelevant nodes and edges.
	Second, we propose an elementary-to-advanced curriculum pre-training strategy to fully utilize useful information in HIN
	for recommendation task.
	We consider that the node/edge/path-level semantics is essential but far from recommendation task, while the descriptive information of the whole subgraph is directly helpful for capturing user-item correlation.
	Inspired by human beings` learning process of difficult problem, we naturally categorize the two types of features into basic and advanced courses for modeling user-item interaction, which provides a smooth inside-out understanding of the user-item interaction subgraph.
	To our knowledge, it is the first time that curriculum pre-training strategy is utilized to bridge the semantic gap between the HIN and recommendation task.}

\section{Experiment}
\label{sec-experiment}
In this section, we first set up the experiments, and then present the results and analysis.

\subsection{Experimental Setup}
\subsubsection{Datasets}
We conduct experiments on three widely-used datasets from different domains, namely Movielens\footnote{\url{https://grouplens.org/datasets/movielens/100k/}}, Amazon\footnote{\url{http://jmcauley.ucsd.edu/data/amazon/links.html}} and Yelp\footnote{\url{https://www.yelp.com/dataset}}, where movies, products and businesses are considered as items for recommendation, respectively. 
We treat a rating as an interaction record, indicating whether a user has rated an item or not.
For reproducible comparison, we reuse the preprocessed results and the selected meta-paths released in \cite{DBLP:conf/cikm/HuSZY18}\footnote{\scriptsize{Accessible via the link: https://github.com/librahu/HIN-Datasets-for-Recommendation-and-Network-Embedding}}. 
The detailed statistics of these datasets after preprocessing are summarized in Table~\ref{tab:datasets}, where we report the statistics by different edge relations. 
The first row of each dataset corresponds to the number of users, items and interactions, while the other rows correspond to the statistics of other relations. The selected meta-paths for each dataset are in the last column.

\begin{table}
		\caption{Basic statistics of the three datasets. }
		\label{tab:datasets}
		\setlength{\tabcolsep}{0.5mm}
		\centering
		\begin{tabular}{llrrrc}
			\toprule
			Datasets & Relations & \#Type A &\#Type B &\#A-B &Metapath\\
			\midrule
			\multirow{4} * {Movielens}
			&\underline{User-Movie} &943 &1,682 &100,000 &UMUM\\
			&Movie-Movie &1,682 &1,682 &82,798 &UMMM\\
			&User-Occupation &943 &21 &943 &UOUM\\ 
			&Movie-Genre &1,682 &18 &2,861 &UMGM\\
			\midrule
			\multirow{4} * {Amazon}
			&\underline{User-Product} &3,584 &2,753 &50,903 &UPUP\\
			&Product-View &2,753 &3,857 &5,694 &UPVP\\
			&Product-Brand &2,753 &334 &2,753 &UPBP\\
			&Product-Category &2,753 &22 &5,508 &UPCP\\
			
			\midrule
			\multirow{4} * {Yelp}
			&\underline{User-Business} &16,239 &14,284 &198,397 &UBUB\\
			&User-User &16,239 &16,239 &158,590 &UUUB\\
			&Business-City &14,284 &47 &14,267 &UBCiB\\
			&Business-Category &14,284 &511 &40,009 &UBCaB\\
			\bottomrule
		\end{tabular}
	\end{table}

\subsubsection{Evaluation Metrics}
We use two commonly used metrics to evaluate the performance of our proposed model.

$\bullet${Hit Rate}: Hit rate (HR) measures the percentage that recommended items contain at least one correct item interacted by the user, which  does
not consider the actual rank of the items and has been widely used in previous works~\cite{DBLP:conf/icdm/Rendle10,DBLP:conf/kdd/Lu0S20}.

\begin{equation}\label{hr-metric}
\text{HR@}k = \frac{1}{|\mathcal{U}|}\sum_{u \in \mathcal{U}}\mathbb{I}(|\hat{\mathcal{I}}_{u, k} \cap \mathcal{I}_{u}|>0),
\end{equation}
where $\hat{\mathcal{I}}_{u, k}$ denotes the set of top-$k$ recommended items for
user $u$ and $\mathcal{I}_{u}$ is the set of testing items for user $u$, and $\mathbb{I}(\cdot)$ is an indicator function.

$\bullet${Normalized Discounted Cumulative Gain}: Normalized Discounted Cumulative Gain (NDCG) takes the positions of
correct recommended items into consideration, which is important in settings where the order of
recommendations matters.
\begin{align}\label{ndcg-metric}
\text{NDCG@}k &= \frac{\text{DCG@}k}{\text{iDCG}}, \nonumber \\ 
\text{DCG@}k &= \frac{1}{|\mathcal{U}|}\sum_{u \in \mathcal{U}}\sum^k_{j=1}\frac{\mathbb{I}(\hat{\mathcal{I}}_{u, j} \in \mathcal{I}_{u})}{\log_2(j+1)}
\end{align}
where $\hat{\mathcal{I}}_{u, j}$ denotes the $j$-th recommended item for the user $u$, 
u, and IDCG denots the ideal discounted cumulative gain, which is a normalization constant and is the maximum possible value of DCG@$k$.

We report results on HR@\{10, 20\} and NDCG@\{10, 20\}.
Following~\cite{DBLP:conf/aaai/RenCLR0R19}, we apply the \textit{leave-one-out} strategy for evaluation. Concretely, for each user, we randomly hold out an interaction record as the test set, another interaction record as the validation set and the remaining data is used for training. Since it is time-consuming to rank all items for every user during evaluation, we pair the ground-truth item with 1000 randomly sampled negative items that the user has not interacted with. We calculate all metrics according to the ranking of the items and report the average score over all test users.

\subsubsection{Baselines} 
We consider the following baselines:
\begin{itemize}
	\item BPR~\cite{DBLP:conf/uai/RendleFGS09} is a classic personalized ranking algorithm that
optimizes the pairwise ranking loss function of latent factor model
with implicit feedback via stochastic gradient descent.
	\item FM~\cite{DBLP:conf/icdm/Rendle10} utilizes a generic factorization machine to model the pairwise interactions between different features and further to characterize second order feature interactions.
	\item NCF~\cite{DBLP:conf/www/HeLZNHC17} integrates both generalized matrix factorization and multi-layer perceptron (MLP) to capture user-item interactions, where the MLP is utilized to explore non-linear interactions between the user and item.
	\item NGCF~\cite{DBLP:conf/www/WangJSWYCY19} adopts GNN layers on the user-item
interaction graph, which exploits the user-item graph structure by propagating embeddings on it to refine user and item representations.
	\item LightGCN~\cite{LightGCN} is the state-of-the-art graph neural network based collaborative filtering model which simplifies the design of GCN to make it more concise and appropriate for recommendation.
	\item HGT~\cite{DBLP:conf/www/HuDWS20} introduces node- and edge-type dependent attention mechanism to model heterogeneous graph, which assigns different weights on neighbors during aggregation to capture the interactions among different types of nodes.
    \item HAN ~\cite{DBLP:conf/www/WangJSWYCY19}  treats meta-paths as
virtual edges to connect nodes and utilizes a hierarchical attention mechanism to capture both node-level and semantic-level information.
	\item HERec~\cite{DBLP:journals/tkde/ShiHZY19} adopts a meta-path based random walk to generate meaningful object sequences for network embedding. And then the embeddings are fused in matrix factorization method for recommendation.
	\item LGRec~\cite{DBLP:conf/cikm/HuSZY18} employs a co-attention mechanism to model most informative local neighbor information, and learns effective global relation representations between users and items in HIN for recommendation.
	\item MCRec~\cite{DBLP:conf/kdd/HuSZY18} utilizes convolutional neural network to construct meta-path embeddings and further leverages co-attention mechanism to model interactions among users, items and meta-paths. 
    \item PF-HIN~\cite{DBLP:journals/corr/abs-2007-03184} designs a ranking-based breadth-first search strategy to generate node sequence and utilizes masked node prediction to pre-train the nodes representations.
	\item GCC~\cite{DBLP:conf/kdd/QiuCDZYDWT20} is a recently proposed pre-training method for homogeneous graph via contrastive learning. We fine-tune the pre-trained model released by the authors on our datasets.
	\item Graph-BERT~\cite{DBLP:journals/corr/abs-2001-05140} is a pre-trained graph neural network solely based on attention mechanism without any graph convolution or aggregation operators.
    \item MTRec~\cite{MTRec} introduces a multi-task learning framework for HIN-based recommender systems. It utilize link prediction as an auxiliary task to improve the recommendation performance.
\end{itemize}

Our baselines can  be roughly categorized into four groups:  (1) BPR, FM, NCF, NGCF and LightGCN are classic or neural collaborative filtering methods;
(2) HGT and HAN are specially designed graph neural networks for modeling HIN;
 (3) HERec, LGRec and MCRec extract meta-path based context from HIN, and then model the paths using neural networks;
 (4) PF-HIN, GCC, Graph-BERT and MTRec are pre-training methods that utilize auxiliary supervised signal to pre-train or regularize the model parameters.

\begin{table*}[t!]
	\caption{Parameter settings of all models.}
	\centering
	\small
	\begin{tabular}{|l|| p{10.5cm} |}
		\hline
		Models & Settings\\\hline\hline
		BPR& embedding-size=$64$, 
			learning-rate=$0.01$, batch-size=$256$,
			RMSprop optimizer
		\\
		\hline
		
		FM& embedding-size=$32$, learning-rate=$0.001$,
			batch-size=$256$, Adam optimizer
		\\
		\hline
		
		NCF& factor-num=$8$,   layer-size=$3$, 
			learning-rate=$0.001$, dropout-rate=$0.5$, sample-neg.-num=$4$
			batch-size=$256$, Adam optimizer
		\\
		\hline
		
		HGT& embedding-size=$64$, hidden-size=$64$ 
			learning-rate=$0.001$ 
			batch-size=$256$, Adam optimizer
		\\
		\hline
		
		HAN& embedding-size=$16$, attention-heads=$8$, dropout-rate=$0.5$
			learning-rate=$0.005$, batch-size=$256$,
		 Adam optimizer
		\\
		\hline
		
		HERec& factor-num=$16$, batch-size=$256$
		\\
		\hline
		
		LGRec& neighbor-size=$100$,   embedding-size=$128$, 
			 hidden-size=$128$, learning-rate=$0.001$, batch-size=$256$, 
			dropout-rate=$0.2$, Adam optimizer
		\\
		\hline
		
		MCRec& embedding-size=$64$, learning-rate=$0.001$,
			batch-size=$256$, Adam optimizer
		\\
		\hline
		
		PF-HIN& layer-size=$2$, attention-heads=$2$, hidden-size=$64$ 
			learning-rate=$0.001$, pre-train-epochs=$50$, dropout-rate=$0.5$, Adam optimizer
		\\
		\hline
		GCC & hidden-size=$128$, batch-size=$256$,  
			learning-rate=$0.001$, Adam optimizer
		\\
		\hline
		Graph-BERT& layer-size=$2$, attention-heads=$2$, hidden-size=$32$, k=$7$, 
			learning-rate=$0.001$, dropout-rate=$0.3$, Adam optimizer
		\\
		\hline
		MTRec& hidden-size=$64$ learning-rate=$0.001$, dropout-rate=$0.5$, Adam optimizer
		\\
		\hline
		
		CHEST (Our)& layer-size=$2$, attention-heads=$2$, hidden-size=$128$, batch-size=$256$, mask-node-prob=$0.4$, mask-edge-prob=$0.2$, learning-rate-for-pretrain=$0.001$, mnp-loss-weight=0.4, mep-loss-weight=0.2, mtp-loss-weight=0.4, learning-rate-for-finetune=$0.0001$, Adam optimizer
		\\
		\hline
		
	\end{tabular}	
	\label{tab-parameters-append}
\end{table*}

\begin{table}
\caption{Performance comparison of different methods on HIN-based recommendation. The best and second best results are in bold and underlined fonts respectively. ``$\dagger$'' indicates the statistical significance for $p<0.01$ compared to the best baseline.}
\label{tab:re}
\centering
\small
\begin{tabular}{l|l|cccc}
\hline
Datasets & Models & Hit Ratio@10 & NDCG@10 &Hit Ratio@20 & NDCG@20 \\
\hline
\multirow{16}{*}{Movielens} & BPR$^\heartsuit$     & 0.3383 & 0.1937 & 0.4624 & 0.2249 \\
& FM$^\heartsuit$      & 0.3722 & 0.2144 & 0.5069 & 0.2482 \\
& NCF$^\diamondsuit$   & 0.3807 & 0.2205 & 0.5122 & 0.2541 \\
& NGCF$^\diamondsuit$  & 0.3945 & 0.2264 & 0.5260 & 0.2593 \\
& LightGCN$^\diamondsuit$  & \underline{0.3977} & \underline{0.2367} & \underline{0.5514} & \underline{0.2753} \\
& HGT$^\diamondsuit$  & 0.2630 & 0.1499 & 0.3828 & 0.1801 \\
& HAN$^\diamondsuit$  & 0.3065 & 0.1613 & 0.4358 & 0.1915 \\
& HERec$^\diamondsuit$ &0.1729  &0.1007 &0.2641  &0.1232  \\
& LGRec$^\diamondsuit$ & 0.3754 & 0.2154 & 0.5111 & 0.2495  \\
& MCRec$^\diamondsuit$  &0.3828 & 0.2218& 0.5345 & 0.2601 \\
& PF-HIN$^\heartsuit$ &0.3054  &0.1822 & 0.4422& 0.2175 \\
& GCC$^\diamondsuit$ &0.3860  &0.2219 & 0.5408 & 0.2612 \\
& Graph-BERT$^\diamondsuit$  &0.3712  &0.2200 &0.5270  &0.2591 \\
& MTRec$^\heartsuit$  & 0.3955 & 0.2231 & 0.5440 & 0.2608 \\
& CHEST & \textbf{0.4401$^\dagger$} & \textbf{0.2495$^\dagger$} & \textbf{0.5981$^\dagger$} & \textbf{0.2892$^\dagger$} \\
\hline
\hline
\multirow{16}{*}{Amazon} & BPR$^\heartsuit$ &0.0791 &0.0390 & 0.1300 &0.0518 \\
& FM$^\heartsuit$      &0.0916 &0.0459 &0.1515 &0.0610 \\
& NCF$^\diamondsuit$   &0.0825 &0.0363 &0.1434 &0.0514 \\
& NGCF$^\diamondsuit$  &0.0887 &0.0417 &0.1485 &0.0567 \\
& LightGCN$^\diamondsuit$ &0.1028 &0.0519 &\underline{0.1605} &0.0664 \\
& HGT$^\diamondsuit$  & 0.0496 &0.0286 &0.0684 &0.0334  \\
& HAN$^\diamondsuit$  & 0.0679 &0.0346 &0.1182 & 0.0472 \\
& HERec$^\diamondsuit$ &0.0355  &0.0154 &0.0867 &0.0282  \\
& LGRec$^\diamondsuit$ & 0.0616 &0.0304 & 0.0969 &0.0393  \\
& MCRec$^\diamondsuit$ & 0.0951 &0.0552 &0.1277 &0.0634 \\
& PF-HIN$^\heartsuit$ &0.0243  &0.0129 & 0.0506 &0.0195 \\
& GCC$^\diamondsuit$ &0.0715  &0.0326 &0.1206 &0.0449 \\
& Graph-BERT$^\diamondsuit$  &0.0893  &0.0422 &0.1511 &0.0577 \\
& MTRec$^\heartsuit$  & \underline{0.1062} & \underline{0.0618} &0.1412 &\underline{0.0706} \\
& CHEST & \textbf{0.1204$^\dagger$} & \textbf{0.0703$^\dagger$} & \textbf{0.1765$^\dagger$}& \textbf{0.0844$^\dagger$} \\
\hline
\hline
\multirow{16}{*}{Yelp} & BPR$^\heartsuit$ & 0.1362 & 0.0717 & 0.2069 & 0.0899 \\
& FM$^\diamondsuit$ & 0.1777 & 0.1031 & 0.2605 & 0.1239 \\
& NCF$^\diamondsuit$ & 0.1451 & 0.0772 & 0.2212 & 0.0963 \\
& NGCF$^\diamondsuit$ & 0.1586 & 0.0851 & 0.2399 & 0.1055 \\
& LightGCN$^\diamondsuit$ & 0.1658 & 0.0895 & 0.2530 & 0.1115 \\
& HGT$^\diamondsuit$  & 0.1689 & 0.0956 & 0.2420 & 0.1139 \\
& HAN$^\diamondsuit$  & 0.1759 & 0.1025 & 0.2532 & 0.1219  \\
& HERec$^\diamondsuit$ &0.1768 &0.0956 &0.2518 & 0.1146  \\
& LGRec$^\diamondsuit$ & 0.1908 & 0.1076 & 0.2762 & 0.1292  \\
& MCRec$^\diamondsuit$  & 0.2193  & 0.1310 & 0.3064 & 0.1528 \\
& PF-HIN$^\heartsuit$ &0.1889 &0.1071 &0.2804  &0.1301  \\
& GCC$^\diamondsuit$ &0.1956 &0.1086 &0.3006  &0.1349  \\
& Graph-BERT$^\diamondsuit$ &0.1894 &0.1043 &0.2865 & 0.1286  \\
& MTRec$^\heartsuit$  & \underline{0.2245} & \underline{0.1332} & \underline{0.3142} &\underline{0.1556} \\
& CHEST & \textbf{0.2441$^\dagger$} & \textbf{0.1441$^\dagger$} & \textbf{0.3630$^\dagger$} & \textbf{0.1711$^\dagger$}  \\
\hline
\end{tabular}

\end{table}

\subsubsection{Implementation Details} 
To compare the performance of these methods,  we either adopt the suggested parameter settings from original papers, or optimize the model performance on the validation set. 
In Table~\ref{tab:re}, models with ``$\diamondsuit$'' are implemented by provided source code while those with ``$\heartsuit$'' are implemented by ourselves. 
For all methods that use meta-paths, we use the same meta-paths as shown in the last column of Table~\ref{tab:datasets} 
and sample five path instances for each meta-path. 
For MCRec, we also pre-learned the latent vectors for nodes to initialize parameters as the authors suggested. 


\ignore{
We take edge type as the \emph{meta-relation} for HGT. The meta-paths selected for those HIN based methods are in the last column of the Table~\ref{tab:datasets}. Note that we adjust the meta-paths type for HERec and HAN which utilize meta-paths with the same starting node and ending node type (\ie{UOU, MGM}).
The number of sampled path instances for each meta-path is 5. Note that we use the same sampled path instances for MCRec, MTRec and our model, and only MCRec utilizes the pre-trained features. 
}

In our model, we use two self-attention blocks each with two attention heads, and set the embedding size as 64. 
In the pre-training stage, the mask proportions of nodes and edges are set as 0.4 and 0.2, and the weights for the three pre-training losses in the elementary course (\ie MNP, MEP and MTP) are set as 0.4, 0.2, and 0.4, and the softmax temperature in the advanced course is set to 1. We use the Adam optimizer~\cite{DBLP:journals/corr/KingmaB14} with learning rates of 0.001 for pre-training and fine-tuning stages.
For the baselines, all the models have some parameters to tune. We either follow the reported optimal parameter settings or optimize each model separately using the validation set. We report the parameter setting used throughout the experiments in Table~\ref{tab-parameters-append}.

\emph{The code and dataset will be available after the review period.}

\subsection{Performance Comparison}

\ignore{
\begin{table*}[t!]
		\caption{Performance comparison of different methods on HIN-based recommendation. HR and ND denote Hit Ratio and NDCG metrics. Models with ``$\diamondsuit$'' are implemented by provided source code while those with ``$\heartsuit$'' are implemented by ourselves. The best and second best results are in bold and underlined fonts respectively. ``$\dagger$'' indicates the statistical significance for $p<0.01$ compared to the best baseline.}
		\label{tab:re}
		\centering
		\footnotesize
		\setlength{\tabcolsep}{0.38mm}{
			\begin{tabular}{l|cccc | cccc|cccc}
				\hline
				\multirow{2}{*}{Models}&\multicolumn{4}{c|}{Movielens}&\multicolumn{4}{c|}{Amazon}&\multicolumn{4}{c}{Yelp}\\
				\cline{2-13}
				&HR@10 & ND@10 &HR@20 & ND@20& HR@10 & ND@10 &HR@20    &ND@20 &HR@10 & ND@10 &HR@20 & ND@20\\
				\hline
				BPR$^\heartsuit$     & 0.3383 & 0.1937 & 0.4624 & 0.2249 &0.0791 &0.0390 & 0.1300 &0.0518& 0.1362 & 0.0717 & 0.2069 & 0.0899 \\
				FM$^\heartsuit$      & 0.3722 & 0.2144 & 0.5069 & 0.2482 &0.0916 &0.0459 &0.1515 &0.0610& 0.1777 & 0.1031 & 0.2605 & 0.1239 \\
				NCF$^\diamondsuit$   & 0.3807 & 0.2205 & 0.5122 & 0.2541 &0.0825 &0.0363 &0.1434 &0.0514& 0.1451 & 0.0772 & 0.2212 & 0.0963 \\
				NGCF$^\diamondsuit$  & 0.3945 & 0.2264 & 0.5260 & 0.2593 &0.0887 &0.0417 &0.1485 &0.0567& 0.1586 & 0.0851 & 0.2399 & 0.1055 \\
				LightGCN$^\diamondsuit$  & \underline{0.3977} & \underline{0.2367} & \underline{0.5514} & \underline{0.2753} &0.1028 &0.0519 &\underline{0.1605} &0.0664& 0.1658 & 0.0895 & 0.2530 & 0.1115 \\
				\hline
				HGT$^\diamondsuit$  & 0.2630 & 0.1499 & 0.3828 & 0.1801 & 0.0496 &0.0286 &0.0684 &0.0334 & 0.1689 & 0.0956 & 0.2420 & 0.1139 \\
				HAN$^\diamondsuit$  & 0.3065 & 0.1613 & 0.4358 & 0.1915 & 0.0679 &0.0346 &0.1182 & 0.0472& 0.1759 & 0.1025 & 0.2532 & 0.1219  \\
				\hline
				HERec$^\diamondsuit$ &0.1729  &0.1007 &0.2641  &0.1232 &0.0355  &0.0154 &0.0867 &0.0282&0.1768 &0.0956 &0.2518 & 0.1146  \\
				LGRec$^\diamondsuit$ & 0.3754 & 0.2154 & 0.5111 & 0.2495& 0.0616 &0.0304 & 0.0969 &0.0393& 0.1908 & 0.1076 & 0.2762 & 0.1292  \\
				MCRec$^\diamondsuit$  &0.3828 & 0.2218& 0.5345 & 0.2601 & 0.0951 &0.0552 &0.1277 &0.0634& 0.2193  & 0.1310 & 0.3064 & 0.1528 \\
				\hline
				PF-HIN$^\heartsuit$ &0.3054  &0.1822 & 0.4422& 0.2175 &0.0243  &0.0129 & 0.0506 &0.0195&0.1889 &0.1071 &0.2804  &0.1301  \\
				GCC$^\diamondsuit$ &0.3860  &0.2219 & 0.5408 & 0.2612 &0.0715  &0.0326 &0.1206 &0.0449&0.1956 &0.1086 &0.3006  &0.1349  \\
				Graph-BERT$^\diamondsuit$  &0.3712  &0.2200 &0.5270  &0.2591&0.0893  &0.0422 &0.1511 &0.0577
				&0.1894 &0.1043 &0.2865 & 0.1286  \\
				MTRec$^\heartsuit$  & 0.3955 & 0.2231 & 0.5440 & 0.2608 & \underline{0.1062} & \underline{0.0618} &0.1412 &\underline{0.0706}& \underline{0.2245} & \underline{0.1332} & \underline{0.3142} &\underline{0.1556} \\
				\hline
				CHEST$_{Base}$ &0.3902  &0.2215 &0.5461  &0.2608 &0.1154  &0.0673 &0.1604 &0.0796&0.2279 &0.1302 &0.3286 &0.1554  \\
				
				CHEST & \textbf{0.4401$^\dagger$} & \textbf{0.2495$^\dagger$} & \textbf{0.5981$^\dagger$} & \textbf{0.2892$^\dagger$} & \textbf{0.1204$^\dagger$} & \textbf{0.0703$^\dagger$} & \textbf{0.1765$^\dagger$}& \textbf{0.0844$^\dagger$}& \textbf{0.2441$^\dagger$} & \textbf{0.1441$^\dagger$} & \textbf{0.3630$^\dagger$} & \textbf{0.1711$^\dagger$}  \\
				\hline
			\end{tabular}
		}
\end{table*}}

Table~\ref{tab:re} presents the performance comparison of different methods on the recommendation task. 

As we can see, for five classic recommendation baselines, FM performs better than BPR, NCF and NGCF on the more sparse datasets (\ie Amazon and Yelp), because FM can incorporate context features and characterize second-order feature interaction. For the three neural collaborative filtering methods, the performance order is consistent across all datasets, \ie LightGCN$>$NGCF$>$NCF. A possible reason is that LightGCN and NGCF utilize graph neural network to learn high-order interaction in a more effective way.

Second, HAN performs much better than HGT in most cases. A major reason is that HGT requires to set meta-relations, which is slightly different from our setting. However, the two methods are general GNN embedding methods, which may be not aware of the recommendation task and cannot perform better than classic recommender systems.

Third, for meta-path based baselines, the performance order is consistent, \ie MCRec$>$LGRec$>$HERec. Because MCRec samples path instances through ``priority'' based strategy and utilizes pre-learned embeddings to initialize the representations in HIN.


Furthermore, the baselines with  auxiliary supervision signals perform better than those from other categories on average. 
Specially, MTRec performs the best among all the baselines, which has designed a special auxiliary task for improving the recommendation performance. 
It relies on self-attention mechanism to learn the semantics of meta-paths in HIN and jointly optimizes the tasks of both recommendation and link prediction. While, GCC and Graph-BERT perform slightly worse than MTRec, which verifies that homogeneous graph pre-training methods cannot be directly utilized in HIN-based recommendation.



Finally, our model CHEST performs consistently better than all the baselines by a large margin on three datasets.
Different from these baselines,  our heterogeneous subgraphs are specially sampled for user-item interaction, which is tailored to the recommendation task.
Besides, our proposed heterogeneous subgraph Transformer is able to preserve graph-structure and path-level semantics within the subgraph via special composite node embeddings.
We further propose the curriculum pre-training strategy to learn effective representations for utilizing useful information in HIN for recommendation task.
Comparing our approach with all the baseline models, it can be observed that the above strategies are very useful to improve the recommendation performance. 

\subsection{Detailed Analysis}

In this section, we perform a series of detailed analysis on the performance of our model. 

\subsubsection{Ablation Study}
\begin{table}
	\caption{Ablation Study of our approach on composite node embeddings.}
	\label{tab:ab_emb}
	\setlength{\tabcolsep}{0.9mm}
	\centering
	\begin{tabular}{ccccc}
	\toprule
		\multirow{2}*{Methods} &  \multicolumn{2}{c}{Movielens} &
		\multicolumn{2}{c}{Amazon} \\
		\cline{2-5}
		\specialrule{0em}{1pt}{2pt}
		&HR@20 &NDCG@20 &HR@20 &NDCG@20 \\
	\midrule
		CHEST  &\textbf{0.5981} &\textbf{0.2892} & \textbf{0.1765} & \textbf{0.0844} \\
	\midrule
		w/o Node Type &0.5928 &0.2835 &0.1666 &0.0817\\
		 w/o Slot &0.5811 &0.2887 &0.1601 &0.0765  \\
		w/o Precursor &0.5811 &0.2855 &0.1485 &0.0696\\
	\bottomrule
	\end{tabular}
\end{table}

\begin{table}
	\caption{Ablation study of our approach on pre-training tasks (P) and other curriculum settings (C).}
	\label{tab:ab_task}
	\centering
	\setlength{\tabcolsep}{0.9mm}
	\centering
	\begin{tabular}{llcccc}
	\toprule
		& \multirow{2}*{Methods} &  \multicolumn{2}{c}{Movielens} &
		\multicolumn{2}{c}{Amazon} \\
		\cline{3-6} 
 		\specialrule{0em}{1pt}{2pt}
		& &HR@20 &NDCG@20 &HR@20 &NDCG@20 \\
	\midrule
		& CHEST  &\textbf{0.5981} &\textbf{0.2892} & \textbf{0.1765} & \textbf{0.0844} \\
	\midrule
	\multirow{4}*{(P)} 
	    &w/o MTP &0.5938 &0.2830 &0.1703 &0.0790 \\
		&w/o MEP &0.5875 &0.2849 &0.1703 &0.0803 \\
		&w/o MNP &0.5239 &0.2437 &0.1587 &0.0724 \\
		&w/o SCL &0.5896 &0.2859 &0.1715 &0.0807 \\
		\hline
		\multirow{2}*{(C)} 
		&Multi-task &0.5589 &0.2715 &0.1640 &0.0819 \\
		&Reverse Courses & 0.5758 & 0.2823 & 0.1700 & 0.0829 \\
	\bottomrule
	\end{tabular}
\end{table}


In our proposed CHEST, we have incorporated four types of node embeddings and design curriculum pre-training strategy for HIN-based recommendation. In this part, we examine the effectiveness of these proposed techniques or components on the model performance.
We conduct the ablation study on Movielens and Amazon datasets, and adopt HR@20 and NDCG@20 as evaluation metrics. 

We first analyze the contribution of the composite embeddings. Besides node ID embeddings, we introduce node type embedding, slot embedding and precursor embedding to preserve the semantics of interaction-specific heterogeneous subgraphs in multi-slot sequence representations. The results after embedding ablation (ID embedding is reserved in all cases) are shown in Table~\ref{tab:ab_emb}.   As we can see, all the embeddings are useful to improve the model performance. Specially, the precursor embedding seems more important than the other two types of embeddings, since it can preserve the graph-structure semantics within the subgraph.

Next, we continue to conduct the ablation study to analyze the contribution of each pre-training task and other curriculum settings. 
As can be seen in Table~\ref{tab:ab_task}, the performance drops when we remove one of the pre-training tasks, which shows that the above tasks are all beneficial to our model.
Among them, the MNP (Masked Node Prediction) is more important than other pre-training tasks. One possible reason is that the correlations between the node and its surrounding context are important for recommendation task.
Under the ``\emph{Multi-task}'' setting, we pre-train the model on four pre-training tasks via multi-task learning, and the performance drops compared to the curriculum learning paradigm.
The ``\emph{Reverse Courses}'' setting means reversing the learning order of the elementary course and the advanced course, which decreases the recommendation performance. These findings verify the rationality of our elementary-to-advanced curriculum learning setting. 


\subsubsection{Subgraph Construction}
\begin{figure}[t!]
    \centering
    \begin{subfigure}[b]{0.35\linewidth}
        \centering
        \includegraphics[width=\textwidth]{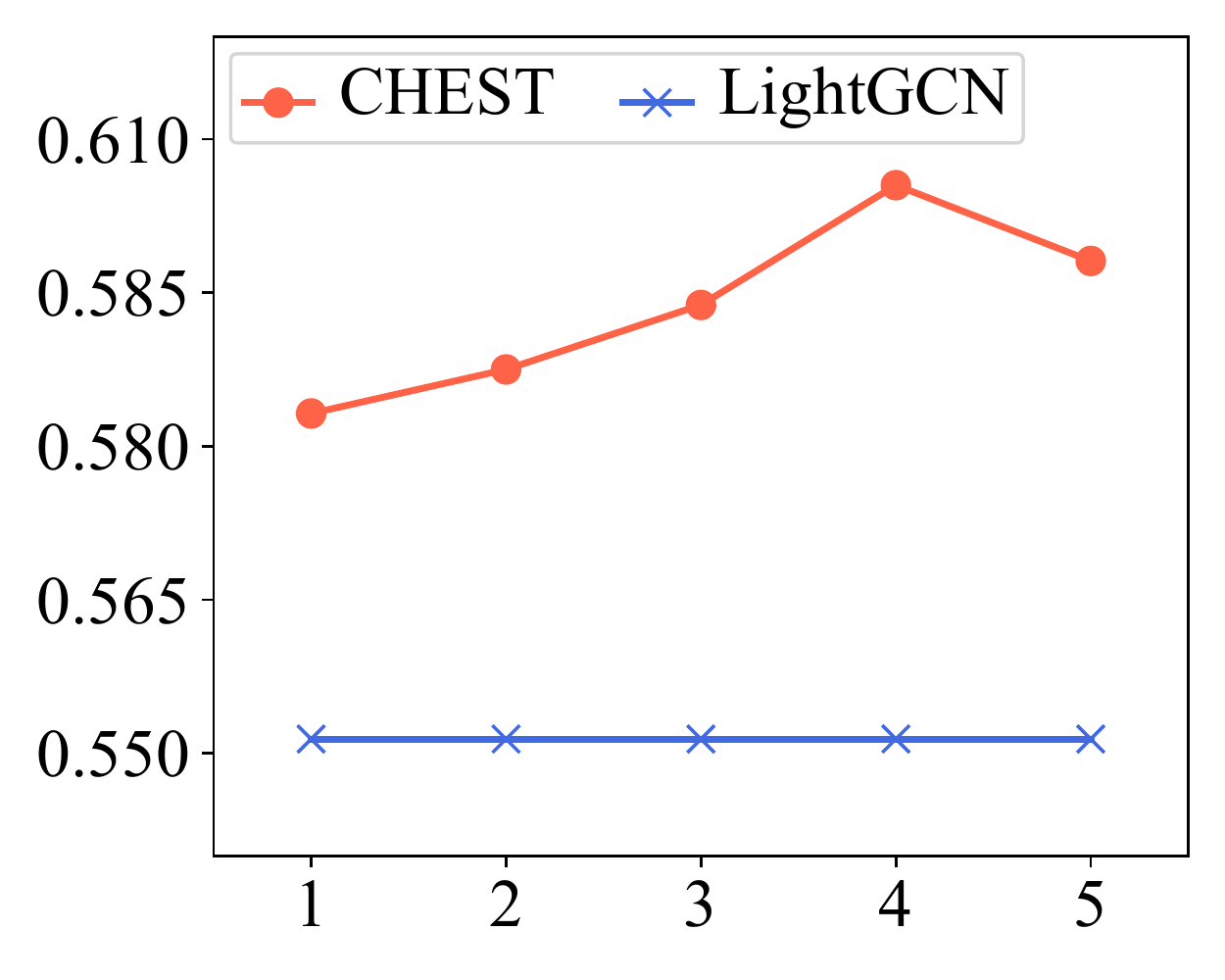}
        \caption{Varying the \#path $K$.}
        \label{ml-num-hr}
    \end{subfigure}
    \begin{subfigure}[b]{0.35\linewidth}
        \centering
        \includegraphics[width=\textwidth]{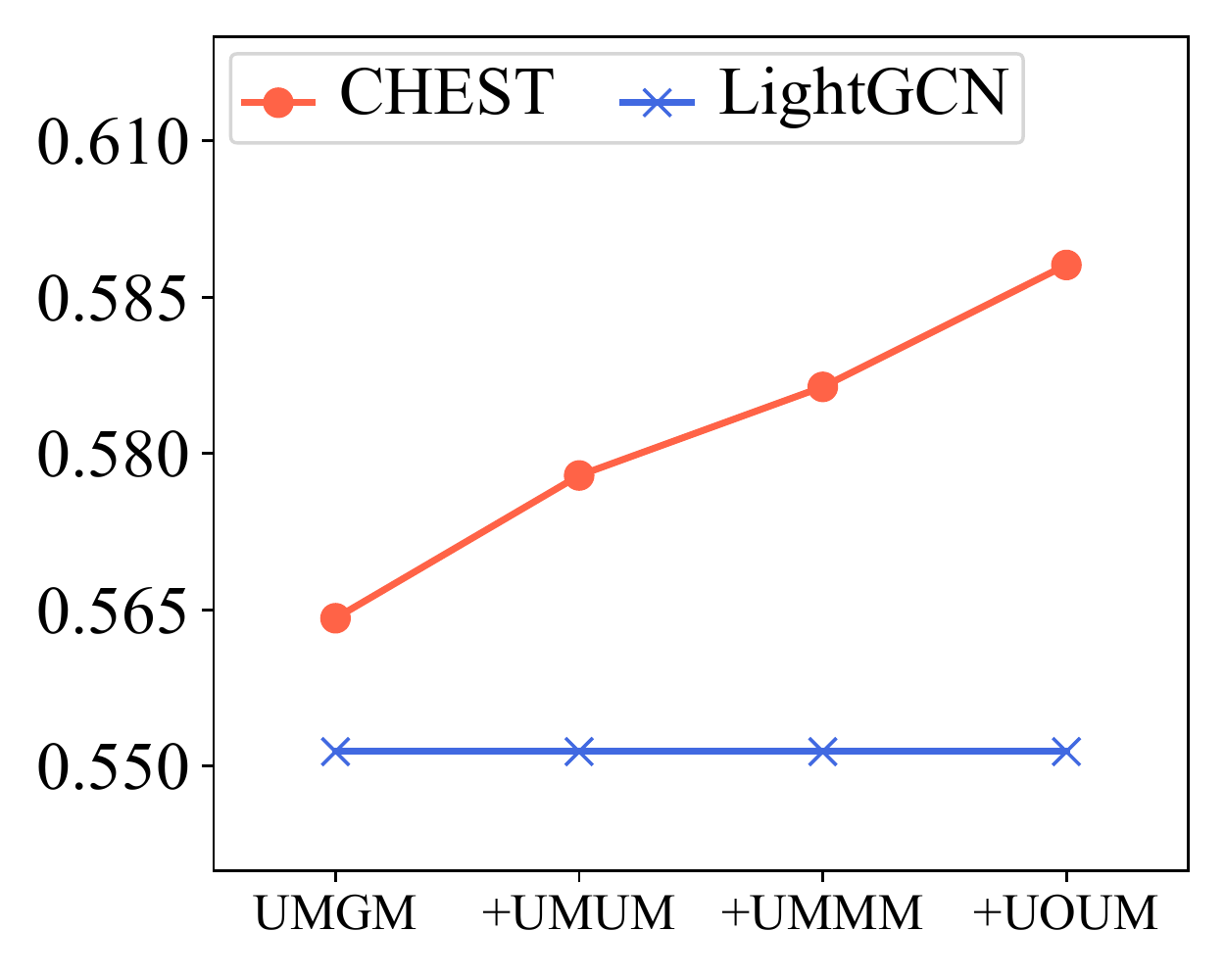}
        \caption{Varying the meta-path types.}
        \label{ml-type-hr}
    \end{subfigure}
    \caption{Performance (HR@20) comparison \emph{w.r.t.} different \#path $K$ and meta-paths types on Movielens dataset.}
    \vspace{-0.2cm}
\label{fig-cons}
\end{figure}
To construct the interaction-specific heterogeneous subgraph, we keep top-$K$ path instances with the highest average similarities for each meta-path. We study the effectiveness of different $K$ on the model performance. As we can see in the Figure~\ref{fig-cons}(a), CHEST achieves good results using only one path instance for each meta-path, which indicates that ``\emph{priority}''-based strategy is able to sample high-quality path instances. But when the $K$ is too large, the results drop a bit. One possible reason is that we may introduce some noisy path instances into the subgraph.

We also investigate the influence of different meta-paths on the recommendation performance by gradually incorporating meta-paths into the subgraph.
As shown in Figure~\ref{fig-cons}(b), the performance of CHEST consistently improves with the incorporation of more meta-paths. The reason is that different meta-paths can introduce different aspects of information for modeling user-item interaction.
\subsubsection{Parameter Tuning}

\begin{figure}[t!]
    \centering
    \begin{subfigure}[b]{0.35\linewidth}
        \centering
        \includegraphics[width=\textwidth]{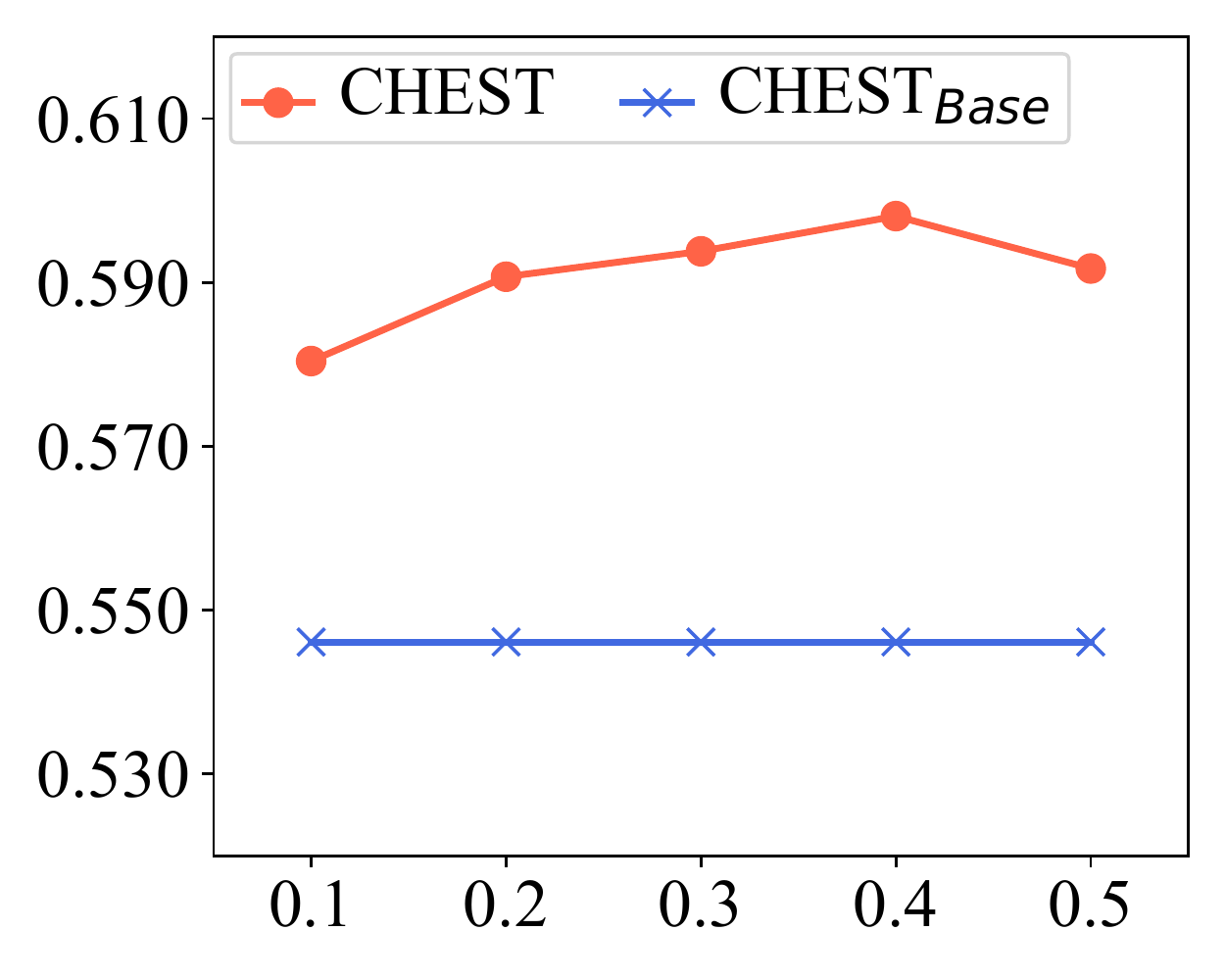}
        \caption{Varying the mask proportion.}
        \label{ml-mask-hr}
    \end{subfigure}
    \begin{subfigure}[b]{0.35\linewidth}
        \centering
        \includegraphics[width=\textwidth]{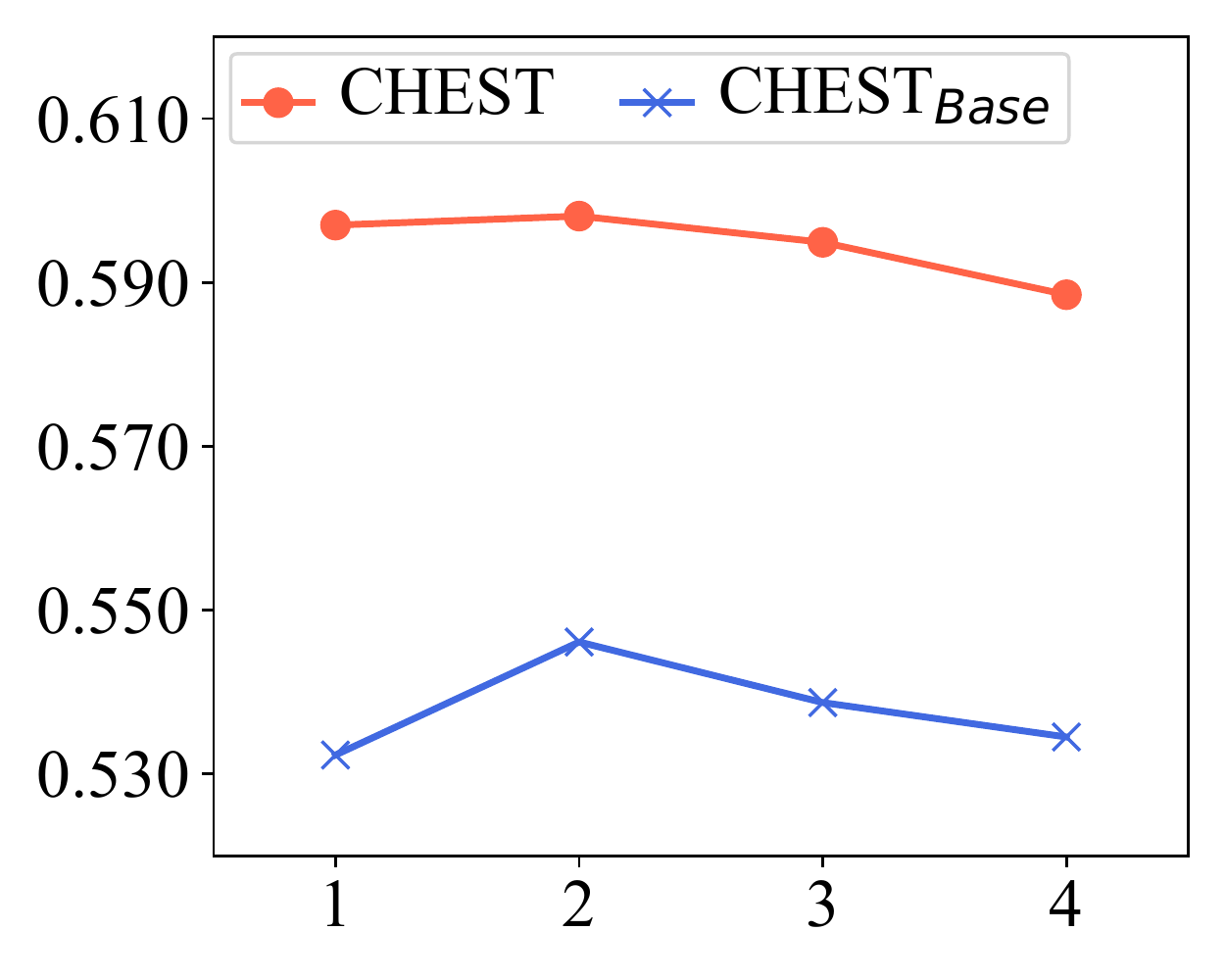}
        \caption{Varying the Trans. layer.}
        \label{ml-mask-hr}
    \end{subfigure}
    \caption{Performance (HR@20) tuning \emph{w.r.t.} different mask proportion and Transformer layers on Movielens dataset.}
    \vspace{-0.2cm}
\label{fig-tuning}
\end{figure}

Our models include a few parameters to tune. Here, we report the tuning results (HR@20) of two parameters on Movielens datasets, \ie the node masked proportion and the number of Transformer layers. The cases on other datasets or metrics are similar and omitted. 

As shown in Figure~\ref{fig-tuning}(a), we can see that CHEST achieves the best performance when the mask proportion is set to 0.4 for Movielens dataset. It indicates that the mask proportion cannot be set too small or too large.
Besides, Figure~\ref{fig-tuning}(b) shows that CHEST achieves the best performance when the layer number is set to 2 for Movielens dataset. With two self-attention layers, our approach can efficiently learn effective information from HIN for recommendation.
It is seen that CHEST is consistently better than CHEST$_{Base}$, which verifies the effectiveness of our curriculum pre-training strategy.

\subsubsection{Data Sparsity}
\begin{figure}[t!]
    \centering
    \includegraphics[width=0.9\linewidth]{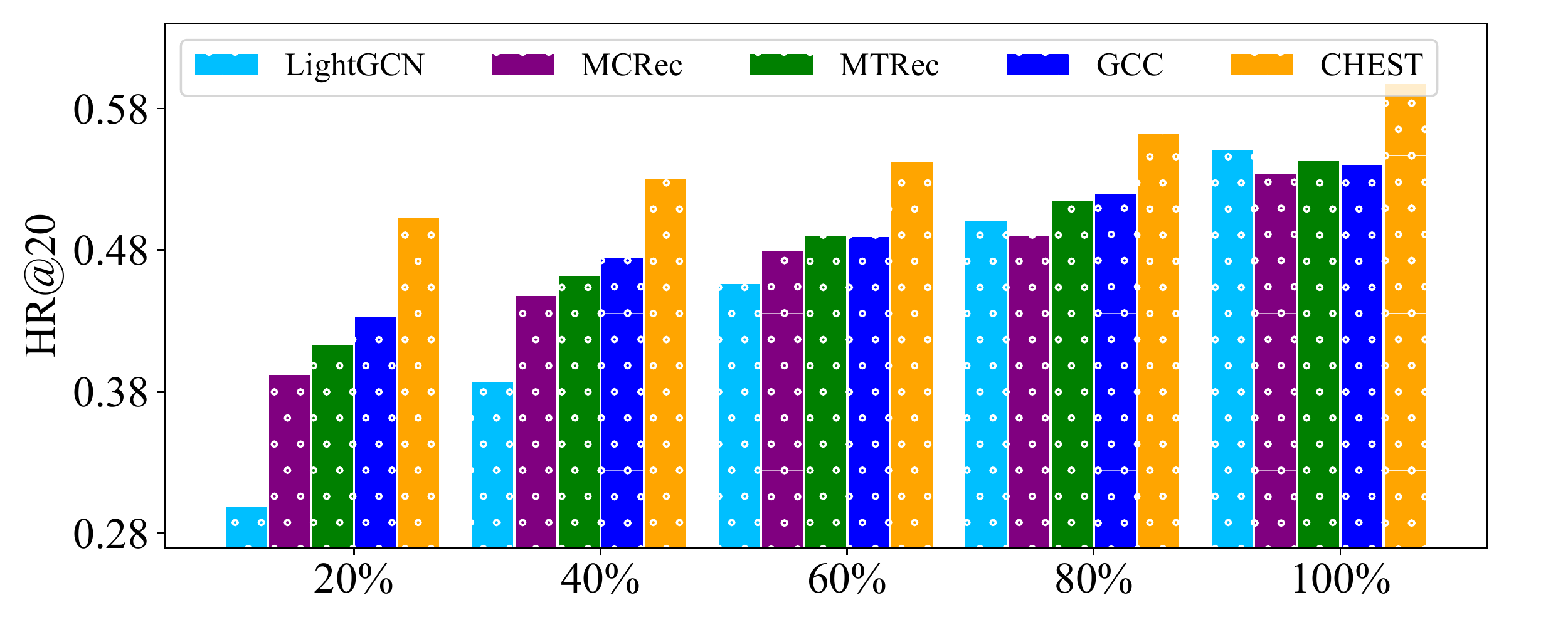}
    \caption{Performance comparison \emph{w.r.t.} different sparsity levels on Movielens dataset.}
    \label{fig-data-amount}
\end{figure}

Recommender systems usually require a considerable amount of training data, thus they are likely to suffer from data sparsity in practice. 
This issue can be alleviated by our method because the proposed curriculum pre-training strategy can leverage intrinsic data correlations from input as auxiliary supervision signals. 
We simulate the data sparsity scenarios by using different proportions of the full training dataset, \ie 20\%, 40\%, 60\%, 80\%, and 100\%. 

Figure~\ref{fig-data-amount} shows the results of data sparsity analysis on Movielens dataset. As we can see, the performance substantially drops when less training data is used.
While, CHEST, GCC and MTRec are consistently better than other methods in data sparsity scenarios, especially in an extreme sparsity level (20\%). It is because these methods utilize auxiliary supervised signals to enhance the data representations or initialize the model parameters. 
Among them, CHEST achieves the best performance since it utilizes an elementary-to-advanced training process to learn effective representations tailored to user-item interactions. 
\ignore{. It shows that
CHEST is able to make better use of the data with the pre-training method, which alleviates the influence of data sparsity problem in recommendation systems to some extent.
}

\subsection{Qualitative Analysis}
\begin{figure}[t!]
    \centering
    \includegraphics[width=.75\linewidth]{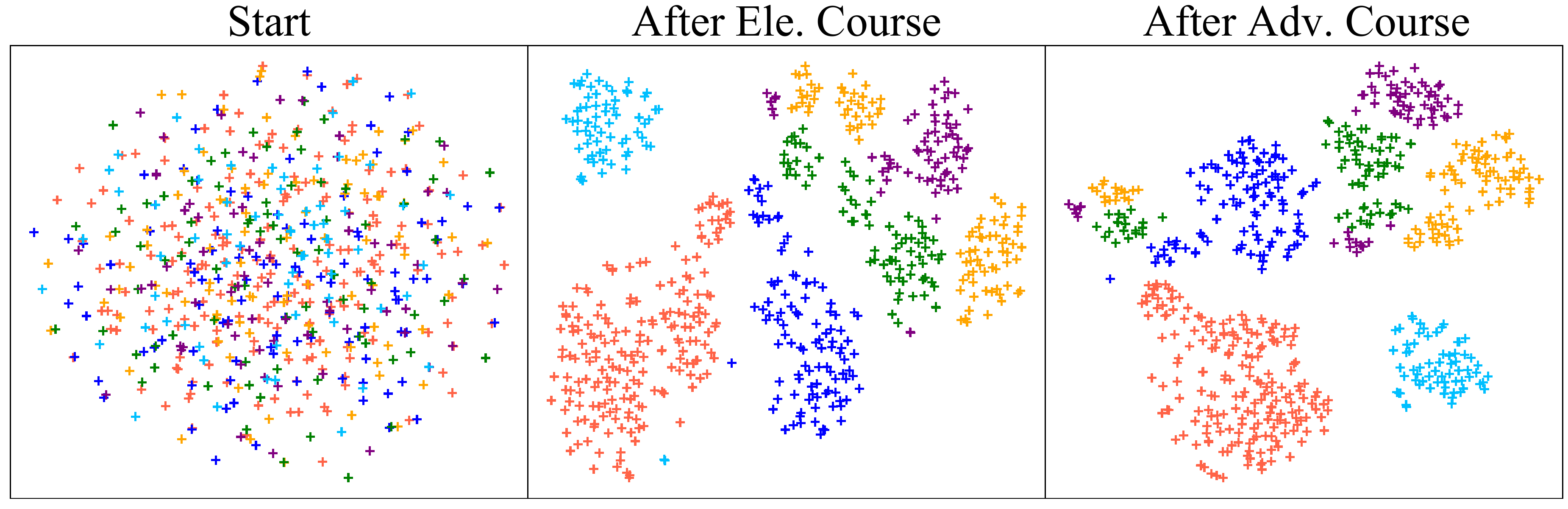}
    \caption{Visualization of the learned user embeddings \emph{w.r.t.} different pre-training courses on Movielens dataset. Colors correspond to the occupations of the users.}
    \label{fig-user-emb}
\end{figure}
\begin{figure}[t!]
    \centering
    \includegraphics[width=.75\linewidth]{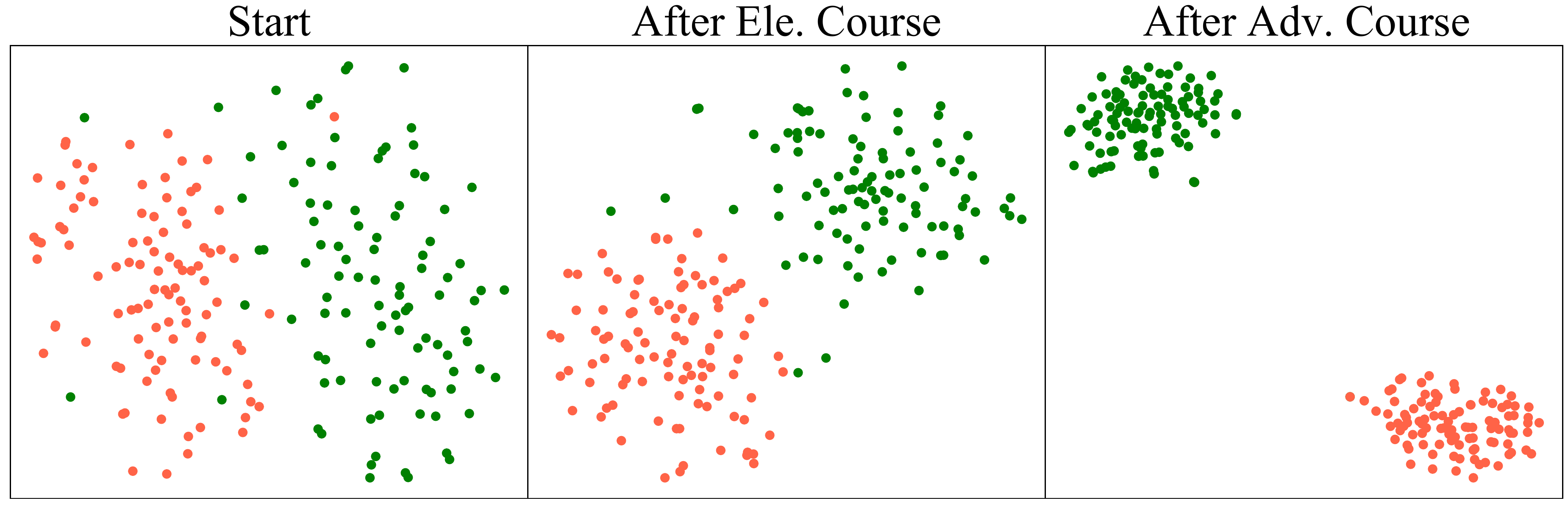}
    \caption{Visualization of the learned subgraph representations \emph{w.r.t.} different pre-training courses on Movielens dataset. Positive samples and negative samples of the same interaction-specific subgraph are in different colors.}
    \label{fig-subgraph-emb}
\end{figure}

The above results have shown the effectiveness of curriculum pre-training strategy for the recommendation task. In this section, we present some qualitative analysis to understand why our approach works. Specially, we present two examples to qualitatively illustrate how the elementary-to-advanced training process improves the learning of data representations. We visualize the two-dimensional projections of learned user embeddings and subgraph representations on Movielens dataset using t-SNE algorithm~\cite{maaten2008visualizing}. 



As shown in Figure~\ref{fig-user-emb}, various colors represent different \emph{occupations} of users in Movielens dataset.
Before pre-training, the representations of users with the same occupations are distributed randomly.
However, after pre-training on the elementary course, our approach derives more coherent clusters corresponding to different occupations. 
After the advanced course, we can see that the produced clusters of user representations are still separated clearly.

In the meantime, Figure~\ref{fig-subgraph-emb} presents the distribution of positive samples (\ie augmented subgraphs) and negative samples of the original interaction-specific subgraph. 
As we can see, after trained on elementary course, the subgraph representations have not been aggregated into coherent clusters. One possible reason is that the elementary course only focuses on the local context information (\eg node, edge and path) by which our model is still unaware of global information of the whole subgraph. 
While these subgraph representations are clearly separated into two clusters (\i.e positive samples and negative samples) after the advanced course.
This phenomenon verifies that the advanced course captures global context information of the subgraph.

The above findings indicate that our curriculum pre-training strategy is able to learn local and global semantics underlying HIN, which can enhance the modeling for user-item interaction.

\section{Conclusion}
\label{sec-conclusion}

In this paper, we proposed a curriculum pre-training based heterogeneous subgraph Transformer (CHEST) for HIN-based recommendation task. 
First, we proposed to use the interaction-specific heterogeneous subgraph to extract sufficient and relevant context information from HIN for each user-item pair.
Then we designed the heterogeneous subgraph Transformer to model the subgraph, in which we incorporated a special composite embedding layer to capture graph-structure and path-level semantics and a self-attentive layer to aggregate the representation for the user-item interaction subgraph.
Furthermore, we designed a curriculum pre-training strategy to gradually learn from both local and global contexts in the subgraph tailored to the recommendation task, in which we devised an elementary-to-advanced learning process to learn effective representations with increasing difficulty levels.
Extensive experiments conducted on three real-world datasets demonstrated the effectiveness of our proposed approach against a number of competitive baselines, especially when only limited training data is available.
	
Currently, we have shown that it is promising to utilize curriculum pre-training technique for HIN-based recommendation.
As future work, we plan to design a more general and effective pre-training strategy for improving neural recommendation algorithms. especially more complex recommendation task
Besides, we will also consider extending our approach to more complex recommendation task, such as multimedia recommendation and conversational recommendation.

\bibliographystyle{ACM-Reference-Format}
\bibliography{sample-base}


\begin{thebibliography}{62}


\ifx \showCODEN    \undefined \def \showCODEN     #1{\unskip}     \fi
\ifx \showDOI      \undefined \def \showDOI       #1{#1}\fi
\ifx \showISBNx    \undefined \def \showISBNx     #1{\unskip}     \fi
\ifx \showISBNxiii \undefined \def \showISBNxiii  #1{\unskip}     \fi
\ifx \showISSN     \undefined \def \showISSN      #1{\unskip}     \fi
\ifx \showLCCN     \undefined \def \showLCCN      #1{\unskip}     \fi
\ifx \shownote     \undefined \def \shownote      #1{#1}          \fi
\ifx \showarticletitle \undefined \def \showarticletitle #1{#1}   \fi
\ifx \showURL      \undefined \def \showURL       {\relax}        \fi
\providecommand\bibfield[2]{#2}
\providecommand\bibinfo[2]{#2}
\providecommand\natexlab[1]{#1}
\providecommand\showeprint[2][]{arXiv:#2}

\bibitem[\protect\citeauthoryear{Bengio, Louradour, Collobert, and
  Weston}{Bengio et~al\mbox{.}}{2009}]%
        {CL_Bengio}
\bibfield{author}{\bibinfo{person}{Yoshua Bengio},
  \bibinfo{person}{J\'{e}r\^{o}me Louradour}, \bibinfo{person}{Ronan
  Collobert}, {and} \bibinfo{person}{Jason Weston}.}
  \bibinfo{year}{2009}\natexlab{}.
\newblock \showarticletitle{Curriculum Learning}. In
  \bibinfo{booktitle}{\emph{Proceedings of the 26th Annual International
  Conference on Machine Learning}} \emph{(\bibinfo{series}{ICML '09})}.
  \bibinfo{publisher}{{ACM}}, \bibinfo{pages}{41–48}.
\newblock


\bibitem[\protect\citeauthoryear{Chen, Kornblith, Norouzi, and Hinton}{Chen
  et~al\mbox{.}}{2020}]%
        {SimCLR}
\bibfield{author}{\bibinfo{person}{Ting Chen}, \bibinfo{person}{Simon
  Kornblith}, \bibinfo{person}{Mohammad Norouzi}, {and}
  \bibinfo{person}{Geoffrey~E. Hinton}.} \bibinfo{year}{2020}\natexlab{}.
\newblock \showarticletitle{A Simple Framework for Contrastive Learning of
  Visual Representations}. In \bibinfo{booktitle}{\emph{{ICML} 2020, 13-18 July
  2020, Virtual Event}}, Vol.~\bibinfo{volume}{119}.
  \bibinfo{publisher}{{PMLR}}, \bibinfo{pages}{1597--1607}.
\newblock


\bibitem[\protect\citeauthoryear{Devlin, Chang, Lee, and Toutanova}{Devlin
  et~al\mbox{.}}{2019}]%
        {DBLP:conf/naacl/DevlinCLT19}
\bibfield{author}{\bibinfo{person}{Jacob Devlin}, \bibinfo{person}{Ming{-}Wei
  Chang}, \bibinfo{person}{Kenton Lee}, {and} \bibinfo{person}{Kristina
  Toutanova}.} \bibinfo{year}{2019}\natexlab{}.
\newblock \showarticletitle{{BERT:} Pre-training of Deep Bidirectional
  Transformers for Language Understanding}. In
  \bibinfo{booktitle}{\emph{{NAACL-HLT} 2019}}. \bibinfo{publisher}{Association
  for Computational Linguistics}, \bibinfo{pages}{4171--4186}.
\newblock


\bibitem[\protect\citeauthoryear{Dong, Chawla, and Swami}{Dong
  et~al\mbox{.}}{2017}]%
        {DBLP:conf/kdd/DongCS17}
\bibfield{author}{\bibinfo{person}{Yuxiao Dong}, \bibinfo{person}{Nitesh~V.
  Chawla}, {and} \bibinfo{person}{Ananthram Swami}.}
  \bibinfo{year}{2017}\natexlab{}.
\newblock \showarticletitle{metapath2vec: Scalable Representation Learning for
  Heterogeneous Networks}. In \bibinfo{booktitle}{\emph{{SIGKDD} 2017}}.
  \bibinfo{publisher}{{ACM}}, \bibinfo{pages}{135--144}.
\newblock


\bibitem[\protect\citeauthoryear{Dong, Hu, Wang, Sun, and Tang}{Dong
  et~al\mbox{.}}{2020}]%
        {HNE_survey_sun_tang}
\bibfield{author}{\bibinfo{person}{Yuxiao Dong}, \bibinfo{person}{Ziniu Hu},
  \bibinfo{person}{Kuansan Wang}, \bibinfo{person}{Yizhou Sun}, {and}
  \bibinfo{person}{Jie Tang}.} \bibinfo{year}{2020}\natexlab{}.
\newblock \showarticletitle{Heterogeneous Network Representation Learning}. In
  \bibinfo{booktitle}{\emph{{IJCAI} 2020}}. \bibinfo{pages}{4861--4867}.
\newblock


\bibitem[\protect\citeauthoryear{Fang, Zhao, and Xiao}{Fang
  et~al\mbox{.}}{2020}]%
        {DBLP:journals/corr/abs-2007-03184}
\bibfield{author}{\bibinfo{person}{Yang Fang}, \bibinfo{person}{Xiang Zhao},
  {and} \bibinfo{person}{Weidong Xiao}.} \bibinfo{year}{2020}\natexlab{}.
\newblock \showarticletitle{Exploring Heterogeneous Information Networks via
  Pre-Training}.
\newblock \bibinfo{journal}{\emph{CoRR}}  \bibinfo{volume}{abs/2007.03184}
  (\bibinfo{year}{2020}).
\newblock


\bibitem[\protect\citeauthoryear{Gong, Tao, Maybank, Liu, Kang, and Yang}{Gong
  et~al\mbox{.}}{2016}]%
        {CL_Object}
\bibfield{author}{\bibinfo{person}{Chen Gong}, \bibinfo{person}{Dacheng Tao},
  \bibinfo{person}{Stephen.~J. Maybank}, \bibinfo{person}{Wei Liu},
  \bibinfo{person}{Guoliang Kang}, {and} \bibinfo{person}{Jie Yang}.}
  \bibinfo{year}{2016}\natexlab{}.
\newblock \showarticletitle{Multi-Modal Curriculum Learning for Semi-Supervised
  Image Classification}.
\newblock \bibinfo{journal}{\emph{IEEE Transactions on Image Processing}}
  \bibinfo{volume}{25}, \bibinfo{number}{7} (\bibinfo{year}{2016}),
  \bibinfo{pages}{3249--3260}.
\newblock
\urldef\tempurl%
\url{https://doi.org/10.1109/TIP.2016.2563981}
\showDOI{\tempurl}


\bibitem[\protect\citeauthoryear{Graves, Bellemare, Menick, Munos, and
  Kavukcuoglu}{Graves et~al\mbox{.}}{2017}]%
        {CL_QA}
\bibfield{author}{\bibinfo{person}{Alex Graves}, \bibinfo{person}{Marc~G.
  Bellemare}, \bibinfo{person}{Jacob Menick}, \bibinfo{person}{R{\'{e}}mi
  Munos}, {and} \bibinfo{person}{Koray Kavukcuoglu}.}
  \bibinfo{year}{2017}\natexlab{}.
\newblock \showarticletitle{Automated Curriculum Learning for Neural Networks}.
  In \bibinfo{booktitle}{\emph{{ICML} 2017, Sydney, NSW, Australia, 6-11 August
  2017}}, Vol.~\bibinfo{volume}{70}. \bibinfo{publisher}{{PMLR}},
  \bibinfo{pages}{1311--1320}.
\newblock


\bibitem[\protect\citeauthoryear{Guo, Tan, Xu, Qin, Chen, and Liu}{Guo
  et~al\mbox{.}}{2020}]%
        {DBLP:conf/aaai/GuoTXQCL20}
\bibfield{author}{\bibinfo{person}{Junliang Guo}, \bibinfo{person}{Xu Tan},
  \bibinfo{person}{Linli Xu}, \bibinfo{person}{Tao Qin},
  \bibinfo{person}{Enhong Chen}, {and} \bibinfo{person}{Tie{-}Yan Liu}.}
  \bibinfo{year}{2020}\natexlab{}.
\newblock \showarticletitle{Fine-Tuning by Curriculum Learning for
  Non-Autoregressive Neural Machine Translation}. In
  \bibinfo{booktitle}{\emph{The Thirty-Fourth {AAAI} Conference on Artificial
  Intelligence, {AAAI} 2020, The Thirty-Second Innovative Applications of
  Artificial Intelligence Conference, {IAAI} 2020, The Tenth {AAAI} Symposium
  on Educational Advances in Artificial Intelligence, {EAAI} 2020, New York,
  NY, USA, February 7-12, 2020}}. \bibinfo{publisher}{{AAAI} Press},
  \bibinfo{pages}{7839--7846}.
\newblock
\urldef\tempurl%
\url{https://aaai.org/ojs/index.php/AAAI/article/view/6289}
\showURL{%
\tempurl}


\bibitem[\protect\citeauthoryear{Guo, Huang, Zhang, Zhuang, Dong, Scott, and
  Huang}{Guo et~al\mbox{.}}{2018}]%
        {CL_CV}
\bibfield{author}{\bibinfo{person}{Sheng Guo}, \bibinfo{person}{Weilin Huang},
  \bibinfo{person}{Haozhi Zhang}, \bibinfo{person}{Chenfan Zhuang},
  \bibinfo{person}{Dengke Dong}, \bibinfo{person}{Matthew~R. Scott}, {and}
  \bibinfo{person}{Dinglong Huang}.} \bibinfo{year}{2018}\natexlab{}.
\newblock \showarticletitle{CurriculumNet: Weakly Supervised Learning from
  Large-Scale Web Images}. In \bibinfo{booktitle}{\emph{{ECCV} 2018, Munich,
  Germany, September 8-14, 2018}}, Vol.~\bibinfo{volume}{11214}.
  \bibinfo{publisher}{Springer}, \bibinfo{pages}{139--154}.
\newblock


\bibitem[\protect\citeauthoryear{He, Fan, Wu, Xie, and Girshick}{He
  et~al\mbox{.}}{2020b}]%
        {MoCo}
\bibfield{author}{\bibinfo{person}{Kaiming He}, \bibinfo{person}{Haoqi Fan},
  \bibinfo{person}{Yuxin Wu}, \bibinfo{person}{Saining Xie}, {and}
  \bibinfo{person}{Ross~B. Girshick}.} \bibinfo{year}{2020}\natexlab{b}.
\newblock \showarticletitle{Momentum Contrast for Unsupervised Visual
  Representation Learning}. In \bibinfo{booktitle}{\emph{{CVPR} 2020, Seattle,
  WA, USA, June 13-19, 2020}}. \bibinfo{publisher}{{IEEE}},
  \bibinfo{pages}{9726--9735}.
\newblock


\bibitem[\protect\citeauthoryear{He, Deng, Wang, Li, Zhang, and Wang}{He
  et~al\mbox{.}}{2020a}]%
        {LightGCN}
\bibfield{author}{\bibinfo{person}{Xiangnan He}, \bibinfo{person}{Kuan Deng},
  \bibinfo{person}{Xiang Wang}, \bibinfo{person}{Yan Li},
  \bibinfo{person}{Yong{-}Dong Zhang}, {and} \bibinfo{person}{Meng Wang}.}
  \bibinfo{year}{2020}\natexlab{a}.
\newblock \showarticletitle{LightGCN: Simplifying and Powering Graph
  Convolution Network for Recommendation}. In \bibinfo{booktitle}{\emph{{SIGIR}
  2020, Virtual Event, China, July 25-30, 2020}}. \bibinfo{publisher}{{ACM}},
  \bibinfo{pages}{639--648}.
\newblock


\bibitem[\protect\citeauthoryear{He, Liao, Zhang, Nie, Hu, and Chua}{He
  et~al\mbox{.}}{2017}]%
        {DBLP:conf/www/HeLZNHC17}
\bibfield{author}{\bibinfo{person}{Xiangnan He}, \bibinfo{person}{Lizi Liao},
  \bibinfo{person}{Hanwang Zhang}, \bibinfo{person}{Liqiang Nie},
  \bibinfo{person}{Xia Hu}, {and} \bibinfo{person}{Tat{-}Seng Chua}.}
  \bibinfo{year}{2017}\natexlab{}.
\newblock \showarticletitle{Neural Collaborative Filtering}. In
  \bibinfo{booktitle}{\emph{{WWW} 2017, Perth, Australia, April 3-7, 2017}}.
  \bibinfo{publisher}{{ACM}}, \bibinfo{pages}{173--182}.
\newblock


\bibitem[\protect\citeauthoryear{Hendrycks, Lee, and Mazeika}{Hendrycks
  et~al\mbox{.}}{2019}]%
        {DBLP:conf/icml/HendrycksLM19}
\bibfield{author}{\bibinfo{person}{Dan Hendrycks}, \bibinfo{person}{Kimin Lee},
  {and} \bibinfo{person}{Mantas Mazeika}.} \bibinfo{year}{2019}\natexlab{}.
\newblock \showarticletitle{Using Pre-Training Can Improve Model Robustness and
  Uncertainty}. In \bibinfo{booktitle}{\emph{{ICML} 2019, 9-15 June 2019, Long
  Beach, California, {USA}}}. \bibinfo{pages}{2712--2721}.
\newblock


\bibitem[\protect\citeauthoryear{Hu, Fang, and Shi}{Hu et~al\mbox{.}}{2019}]%
        {DBLP:conf/kdd/HuFS19}
\bibfield{author}{\bibinfo{person}{Binbin Hu}, \bibinfo{person}{Yuan Fang},
  {and} \bibinfo{person}{Chuan Shi}.} \bibinfo{year}{2019}\natexlab{}.
\newblock \showarticletitle{Adversarial Learning on Heterogeneous Information
  Networks}. In \bibinfo{booktitle}{\emph{{SIGKDD} 2019}}.
  \bibinfo{pages}{120--129}.
\newblock


\bibitem[\protect\citeauthoryear{Hu, Shi, Zhao, and Yang}{Hu
  et~al\mbox{.}}{2018a}]%
        {DBLP:conf/cikm/HuSZY18}
\bibfield{author}{\bibinfo{person}{Binbin Hu}, \bibinfo{person}{Chuan Shi},
  \bibinfo{person}{Wayne~Xin Zhao}, {and} \bibinfo{person}{Tianchi Yang}.}
  \bibinfo{year}{2018}\natexlab{a}.
\newblock \showarticletitle{Local and Global Information Fusion for Top-N
  Recommendation in Heterogeneous Information Network}. In
  \bibinfo{booktitle}{\emph{{CIKM} 2018}}. \bibinfo{pages}{1683--1686}.
\newblock


\bibitem[\protect\citeauthoryear{Hu, Shi, Zhao, and Yu}{Hu
  et~al\mbox{.}}{2018b}]%
        {DBLP:conf/kdd/HuSZY18}
\bibfield{author}{\bibinfo{person}{Binbin Hu}, \bibinfo{person}{Chuan Shi},
  \bibinfo{person}{Wayne~Xin Zhao}, {and} \bibinfo{person}{Philip~S. Yu}.}
  \bibinfo{year}{2018}\natexlab{b}.
\newblock \showarticletitle{Leveraging Meta-path based Context for Top- {N}
  Recommendation with {A} Neural Co-Attention Model}. In
  \bibinfo{booktitle}{\emph{{SIGKDD} 2018}}. \bibinfo{pages}{1531--1540}.
\newblock


\bibitem[\protect\citeauthoryear{Hu, Liu, Gomes, Zitnik, Liang, Pande, and
  Leskovec}{Hu et~al\mbox{.}}{2020c}]%
        {DBLP:conf/iclr/HuLGZLPL20}
\bibfield{author}{\bibinfo{person}{Weihua Hu}, \bibinfo{person}{Bowen Liu},
  \bibinfo{person}{Joseph Gomes}, \bibinfo{person}{Marinka Zitnik},
  \bibinfo{person}{Percy Liang}, \bibinfo{person}{Vijay~S. Pande}, {and}
  \bibinfo{person}{Jure Leskovec}.} \bibinfo{year}{2020}\natexlab{c}.
\newblock \showarticletitle{Strategies for Pre-training Graph Neural Networks}.
  In \bibinfo{booktitle}{\emph{{ICLR} 2020, Addis Ababa, Ethiopia, April 26-30,
  2020}}.
\newblock


\bibitem[\protect\citeauthoryear{Hu, Dong, Wang, Chang, and Sun}{Hu
  et~al\mbox{.}}{2020b}]%
        {DBLP:conf/kdd/HuDWCS20}
\bibfield{author}{\bibinfo{person}{Ziniu Hu}, \bibinfo{person}{Yuxiao Dong},
  \bibinfo{person}{Kuansan Wang}, \bibinfo{person}{Kai{-}Wei Chang}, {and}
  \bibinfo{person}{Yizhou Sun}.} \bibinfo{year}{2020}\natexlab{b}.
\newblock \showarticletitle{{GPT-GNN:} Generative Pre-Training of Graph Neural
  Networks}. In \bibinfo{booktitle}{\emph{{KDD} 2020}}.
  \bibinfo{pages}{1857--1867}.
\newblock


\bibitem[\protect\citeauthoryear{Hu, Dong, Wang, and Sun}{Hu
  et~al\mbox{.}}{2020a}]%
        {DBLP:conf/www/HuDWS20}
\bibfield{author}{\bibinfo{person}{Ziniu Hu}, \bibinfo{person}{Yuxiao Dong},
  \bibinfo{person}{Kuansan Wang}, {and} \bibinfo{person}{Yizhou Sun}.}
  \bibinfo{year}{2020}\natexlab{a}.
\newblock \showarticletitle{Heterogeneous Graph Transformer}. In
  \bibinfo{booktitle}{\emph{{WWW} 2020}}. \bibinfo{publisher}{{ACM} / {IW3C2}},
  \bibinfo{pages}{2704--2710}.
\newblock


\bibitem[\protect\citeauthoryear{Isinkaye, Folajimi, and Ojokoh}{Isinkaye
  et~al\mbox{.}}{2015}]%
        {isinkaye2015recommendation}
\bibfield{author}{\bibinfo{person}{Folasade~Olubusola Isinkaye},
  \bibinfo{person}{YO Folajimi}, {and} \bibinfo{person}{Bolande~Adefowoke
  Ojokoh}.} \bibinfo{year}{2015}\natexlab{}.
\newblock \showarticletitle{Recommendation systems: Principles, methods and
  evaluation}.
\newblock \bibinfo{journal}{\emph{Egyptian Informatics Journal}}
  \bibinfo{volume}{16}, \bibinfo{number}{3} (\bibinfo{year}{2015}),
  \bibinfo{pages}{261--273}.
\newblock


\bibitem[\protect\citeauthoryear{Jin, Qin, Fang, Du, Zhang, Yu, Zhang, and
  Smola}{Jin et~al\mbox{.}}{2020}]%
        {DBLP:conf/kdd/JinQFD00ZS20}
\bibfield{author}{\bibinfo{person}{Jiarui Jin}, \bibinfo{person}{Jiarui Qin},
  \bibinfo{person}{Yuchen Fang}, \bibinfo{person}{Kounianhua Du},
  \bibinfo{person}{Weinan Zhang}, \bibinfo{person}{Yong Yu},
  \bibinfo{person}{Zheng Zhang}, {and} \bibinfo{person}{Alexander~J. Smola}.}
  \bibinfo{year}{2020}\natexlab{}.
\newblock \showarticletitle{An Efficient Neighborhood-based Interaction Model
  for Recommendation on Heterogeneous Graph}. In
  \bibinfo{booktitle}{\emph{{SIGKDD} 2020}}. \bibinfo{pages}{75--84}.
\newblock


\bibitem[\protect\citeauthoryear{Kingma and Ba}{Kingma and Ba}{2015}]%
        {DBLP:journals/corr/KingmaB14}
\bibfield{author}{\bibinfo{person}{Diederik~P. Kingma} {and}
  \bibinfo{person}{Jimmy Ba}.} \bibinfo{year}{2015}\natexlab{}.
\newblock \showarticletitle{Adam: {A} Method for Stochastic Optimization}. In
  \bibinfo{booktitle}{\emph{3rd International Conference on Learning
  Representations, {ICLR} 2015, San Diego, CA, USA, May 7-9, 2015, Conference
  Track Proceedings}}.
\newblock


\bibitem[\protect\citeauthoryear{Kipf and Welling}{Kipf and Welling}{2017}]%
        {DBLP:conf/iclr/KipfW17}
\bibfield{author}{\bibinfo{person}{Thomas~N. Kipf} {and} \bibinfo{person}{Max
  Welling}.} \bibinfo{year}{2017}\natexlab{}.
\newblock \showarticletitle{Semi-Supervised Classification with Graph
  Convolutional Networks}. In \bibinfo{booktitle}{\emph{{ICLR} 2017, Toulon,
  France, April 24-26, 2017, Conference Track Proceedings}}.
\newblock


\bibitem[\protect\citeauthoryear{Kong, de~Masson~d'Autume, Yu, Ling, Dai, and
  Yogatama}{Kong et~al\mbox{.}}{2020}]%
        {DBLP:conf/iclr/KongdYLDY20}
\bibfield{author}{\bibinfo{person}{Lingpeng Kong}, \bibinfo{person}{Cyprien de
  Masson~d'Autume}, \bibinfo{person}{Lei Yu}, \bibinfo{person}{Wang Ling},
  \bibinfo{person}{Zihang Dai}, {and} \bibinfo{person}{Dani Yogatama}.}
  \bibinfo{year}{2020}\natexlab{}.
\newblock \showarticletitle{A Mutual Information Maximization Perspective of
  Language Representation Learning}. In \bibinfo{booktitle}{\emph{8th
  International Conference on Learning Representations, {ICLR} 2020, Addis
  Ababa, Ethiopia, April 26-30, 2020}}. \bibinfo{publisher}{OpenReview.net}.
\newblock


\bibitem[\protect\citeauthoryear{Koren and Bell}{Koren and Bell}{2015}]%
        {DBLP:reference/sp/KorenB15}
\bibfield{author}{\bibinfo{person}{Yehuda Koren} {and}
  \bibinfo{person}{Robert~M. Bell}.} \bibinfo{year}{2015}\natexlab{}.
\newblock \showarticletitle{Advances in Collaborative Filtering}.
\newblock In \bibinfo{booktitle}{\emph{Recommender Systems Handbook}},
  \bibfield{editor}{\bibinfo{person}{Francesco Ricci}, \bibinfo{person}{Lior
  Rokach}, {and} \bibinfo{person}{Bracha Shapira}} (Eds.).
  \bibinfo{publisher}{Springer}, \bibinfo{pages}{77--118}.
\newblock
\urldef\tempurl%
\url{https://doi.org/10.1007/978-1-4899-7637-6\_3}
\showDOI{\tempurl}


\bibitem[\protect\citeauthoryear{Koren, Bell, and Volinsky}{Koren
  et~al\mbox{.}}{2009}]%
        {DBLP:journals/computer/KorenBV09}
\bibfield{author}{\bibinfo{person}{Yehuda Koren}, \bibinfo{person}{Robert~M.
  Bell}, {and} \bibinfo{person}{Chris Volinsky}.}
  \bibinfo{year}{2009}\natexlab{}.
\newblock \showarticletitle{Matrix Factorization Techniques for Recommender
  Systems}.
\newblock \bibinfo{journal}{\emph{Computer}} \bibinfo{volume}{42},
  \bibinfo{number}{8} (\bibinfo{year}{2009}), \bibinfo{pages}{30--37}.
\newblock


\bibitem[\protect\citeauthoryear{Li, Ahmed, Ravi, and Smola}{Li
  et~al\mbox{.}}{2014}]%
        {DBLP:conf/kdd/LiARS14}
\bibfield{author}{\bibinfo{person}{Aaron~Q. Li}, \bibinfo{person}{Amr Ahmed},
  \bibinfo{person}{Sujith Ravi}, {and} \bibinfo{person}{Alexander~J. Smola}.}
  \bibinfo{year}{2014}\natexlab{}.
\newblock \showarticletitle{Reducing the sampling complexity of topic models}.
  In \bibinfo{booktitle}{\emph{{KDD} '14, New York, NY, {USA} - August 24 - 27,
  2014}}. \bibinfo{pages}{891--900}.
\newblock


\bibitem[\protect\citeauthoryear{Li, Wang, Lyu, and Shi}{Li
  et~al\mbox{.}}{2020}]%
        {MTRec}
\bibfield{author}{\bibinfo{person}{Hui Li}, \bibinfo{person}{Yanlin Wang},
  \bibinfo{person}{Ziyu Lyu}, {and} \bibinfo{person}{Jieming Shi}.}
  \bibinfo{year}{2020}\natexlab{}.
\newblock \showarticletitle{Multi-task Learning for Recommendation over
  Heterogeneous Information Network}.
\newblock \bibinfo{journal}{\emph{IEEE Transactions on Knowledge and Data
  Engineering}} (\bibinfo{year}{2020}).
\newblock


\bibitem[\protect\citeauthoryear{Ling, Lyu, and King}{Ling
  et~al\mbox{.}}{2014}]%
        {DBLP:conf/recsys/LingLK14}
\bibfield{author}{\bibinfo{person}{Guang Ling}, \bibinfo{person}{Michael~R.
  Lyu}, {and} \bibinfo{person}{Irwin King}.} \bibinfo{year}{2014}\natexlab{}.
\newblock \showarticletitle{Ratings meet reviews, a combined approach to
  recommend}. In \bibinfo{booktitle}{\emph{Eighth {ACM} Conference on
  Recommender Systems, RecSys '14, Foster City, Silicon Valley, CA, {USA} -
  October 06 - 10, 2014}}, \bibfield{editor}{\bibinfo{person}{Alfred Kobsa},
  \bibinfo{person}{Michelle~X. Zhou}, \bibinfo{person}{Martin Ester}, {and}
  \bibinfo{person}{Yehuda Koren}} (Eds.). \bibinfo{publisher}{{ACM}},
  \bibinfo{pages}{105--112}.
\newblock
\urldef\tempurl%
\url{https://doi.org/10.1145/2645710.2645728}
\showDOI{\tempurl}


\bibitem[\protect\citeauthoryear{Liu, He, Liu, and Zhao}{Liu
  et~al\mbox{.}}{2018}]%
        {CL_NLP}
\bibfield{author}{\bibinfo{person}{Cao Liu}, \bibinfo{person}{Shizhu He},
  \bibinfo{person}{Kang Liu}, {and} \bibinfo{person}{Jun Zhao}.}
  \bibinfo{year}{2018}\natexlab{}.
\newblock \showarticletitle{Curriculum Learning for Natural Answer Generation}.
  In \bibinfo{booktitle}{\emph{{IJCAI} 2018, July 13-19, 2018, Stockholm,
  Sweden}}. \bibinfo{publisher}{ijcai.org}, \bibinfo{pages}{4223--4229}.
\newblock


\bibitem[\protect\citeauthoryear{Liu, Ren, Tan, Zhang, Qin, Zhao, and Liu}{Liu
  et~al\mbox{.}}{2020}]%
        {CL_NLP_task_level}
\bibfield{author}{\bibinfo{person}{Jinglin Liu}, \bibinfo{person}{Yi Ren},
  \bibinfo{person}{Xu Tan}, \bibinfo{person}{Chen Zhang}, \bibinfo{person}{Tao
  Qin}, \bibinfo{person}{Zhou Zhao}, {and} \bibinfo{person}{Tie{-}Yan Liu}.}
  \bibinfo{year}{2020}\natexlab{}.
\newblock \showarticletitle{Task-Level Curriculum Learning for
  Non-Autoregressive Neural Machine Translation}. In
  \bibinfo{booktitle}{\emph{{IJCAI} 2020}}. \bibinfo{pages}{3861--3867}.
\newblock


\bibitem[\protect\citeauthoryear{Lu, Fang, and Shi}{Lu et~al\mbox{.}}{2020}]%
        {DBLP:conf/kdd/Lu0S20}
\bibfield{author}{\bibinfo{person}{Yuanfu Lu}, \bibinfo{person}{Yuan Fang},
  {and} \bibinfo{person}{Chuan Shi}.} \bibinfo{year}{2020}\natexlab{}.
\newblock \showarticletitle{Meta-learning on Heterogeneous Information Networks
  for Cold-start Recommendation}. In \bibinfo{booktitle}{\emph{{KDD} 2020,
  Virtual Event, CA, USA, August 23-27, 2020}}. \bibinfo{pages}{1563--1573}.
\newblock


\bibitem[\protect\citeauthoryear{Luo, Pang, Wang, and Lin}{Luo
  et~al\mbox{.}}{2014}]%
        {DBLP:conf/icdm/LuoPWL14}
\bibfield{author}{\bibinfo{person}{Chen Luo}, \bibinfo{person}{Wei Pang},
  \bibinfo{person}{Zhe Wang}, {and} \bibinfo{person}{Chenghua Lin}.}
  \bibinfo{year}{2014}\natexlab{}.
\newblock \showarticletitle{Hete-CF: Social-Based Collaborative Filtering
  Recommendation Using Heterogeneous Relations}. In
  \bibinfo{booktitle}{\emph{2014 {IEEE} International Conference on Data
  Mining, {ICDM} 2014, Shenzhen, China, December 14-17, 2014}},
  \bibfield{editor}{\bibinfo{person}{Ravi Kumar}, \bibinfo{person}{Hannu
  Toivonen}, \bibinfo{person}{Jian Pei}, \bibinfo{person}{Joshua~Zhexue Huang},
  {and} \bibinfo{person}{Xindong Wu}} (Eds.). \bibinfo{publisher}{{IEEE}
  Computer Society}, \bibinfo{pages}{917--922}.
\newblock
\urldef\tempurl%
\url{https://doi.org/10.1109/ICDM.2014.64}
\showDOI{\tempurl}


\bibitem[\protect\citeauthoryear{Ma, Zhou, Liu, Lyu, and King}{Ma
  et~al\mbox{.}}{2011}]%
        {DBLP:conf/wsdm/MaZLLK11}
\bibfield{author}{\bibinfo{person}{Hao Ma}, \bibinfo{person}{Dengyong Zhou},
  \bibinfo{person}{Chao Liu}, \bibinfo{person}{Michael~R. Lyu}, {and}
  \bibinfo{person}{Irwin King}.} \bibinfo{year}{2011}\natexlab{}.
\newblock \showarticletitle{Recommender systems with social regularization}. In
  \bibinfo{booktitle}{\emph{Proceedings of the Forth International Conference
  on Web Search and Web Data Mining, {WSDM} 2011, Hong Kong, China, February
  9-12, 2011}}, \bibfield{editor}{\bibinfo{person}{Irwin King},
  \bibinfo{person}{Wolfgang Nejdl}, {and} \bibinfo{person}{Hang Li}} (Eds.).
  \bibinfo{publisher}{{ACM}}, \bibinfo{pages}{287--296}.
\newblock
\urldef\tempurl%
\url{https://doi.org/10.1145/1935826.1935877}
\showDOI{\tempurl}


\bibitem[\protect\citeauthoryear{Maaten and Hinton}{Maaten and Hinton}{2008}]%
        {maaten2008visualizing}
\bibfield{author}{\bibinfo{person}{Laurens van~der Maaten} {and}
  \bibinfo{person}{Geoffrey Hinton}.} \bibinfo{year}{2008}\natexlab{}.
\newblock \showarticletitle{Visualizing data using t-SNE}.
\newblock \bibinfo{journal}{\emph{Journal of machine learning research}}
  \bibinfo{volume}{9}, \bibinfo{number}{Nov} (\bibinfo{year}{2008}),
  \bibinfo{pages}{2579--2605}.
\newblock


\bibitem[\protect\citeauthoryear{Pham, Li, Cong, and Zhang}{Pham
  et~al\mbox{.}}{2016}]%
        {DBLP:journals/tkde/PhamLCZ16}
\bibfield{author}{\bibinfo{person}{Tuan{-}Anh~Nguyen Pham},
  \bibinfo{person}{Xutao Li}, \bibinfo{person}{Gao Cong}, {and}
  \bibinfo{person}{Zhenjie Zhang}.} \bibinfo{year}{2016}\natexlab{}.
\newblock \showarticletitle{A General Recommendation Model for Heterogeneous
  Networks}.
\newblock \bibinfo{journal}{\emph{{IEEE} Trans. Knowl. Data Eng.}}
  \bibinfo{volume}{28}, \bibinfo{number}{12} (\bibinfo{year}{2016}),
  \bibinfo{pages}{3140--3153}.
\newblock


\bibitem[\protect\citeauthoryear{Qiu, Chen, Dong, Zhang, Yang, Ding, Wang, and
  Tang}{Qiu et~al\mbox{.}}{2020}]%
        {DBLP:conf/kdd/QiuCDZYDWT20}
\bibfield{author}{\bibinfo{person}{Jiezhong Qiu}, \bibinfo{person}{Qibin Chen},
  \bibinfo{person}{Yuxiao Dong}, \bibinfo{person}{Jing Zhang},
  \bibinfo{person}{Hongxia Yang}, \bibinfo{person}{Ming Ding},
  \bibinfo{person}{Kuansan Wang}, {and} \bibinfo{person}{Jie Tang}.}
  \bibinfo{year}{2020}\natexlab{}.
\newblock \showarticletitle{{GCC:} Graph Contrastive Coding for Graph Neural
  Network Pre-Training}. In \bibinfo{booktitle}{\emph{{KDD} 2020}}.
  \bibinfo{pages}{1150--1160}.
\newblock


\bibitem[\protect\citeauthoryear{Ren, Chen, Li, Ren, Ma, and de~Rijke}{Ren
  et~al\mbox{.}}{2019a}]%
        {DBLP:conf/aaai/RenCLR0R19}
\bibfield{author}{\bibinfo{person}{Pengjie Ren}, \bibinfo{person}{Zhumin Chen},
  \bibinfo{person}{Jing Li}, \bibinfo{person}{Zhaochun Ren},
  \bibinfo{person}{Jun Ma}, {and} \bibinfo{person}{Maarten de Rijke}.}
  \bibinfo{year}{2019}\natexlab{a}.
\newblock \showarticletitle{RepeatNet: {A} Repeat Aware Neural Recommendation
  Machine for Session-Based Recommendation}. In
  \bibinfo{booktitle}{\emph{{AAAI} 2019}}. \bibinfo{pages}{4806--4813}.
\newblock


\bibitem[\protect\citeauthoryear{Ren, Liu, Huang, Dai, Bo, and Zhang}{Ren
  et~al\mbox{.}}{2019b}]%
        {DBLP:journals/corr/abs-1911-08538}
\bibfield{author}{\bibinfo{person}{Yuxiang Ren}, \bibinfo{person}{Bo Liu},
  \bibinfo{person}{Chao Huang}, \bibinfo{person}{Peng Dai},
  \bibinfo{person}{Liefeng Bo}, {and} \bibinfo{person}{Jiawei Zhang}.}
  \bibinfo{year}{2019}\natexlab{b}.
\newblock \showarticletitle{Heterogeneous Deep Graph Infomax}.
\newblock \bibinfo{journal}{\emph{CoRR}}  \bibinfo{volume}{abs/1911.08538}
  (\bibinfo{year}{2019}).
\newblock
\showeprint[arxiv]{1911.08538}
\urldef\tempurl%
\url{http://arxiv.org/abs/1911.08538}
\showURL{%
\tempurl}


\bibitem[\protect\citeauthoryear{Rendle}{Rendle}{2010}]%
        {DBLP:conf/icdm/Rendle10}
\bibfield{author}{\bibinfo{person}{Steffen Rendle}.}
  \bibinfo{year}{2010}\natexlab{}.
\newblock \showarticletitle{Factorization Machines}. In
  \bibinfo{booktitle}{\emph{{ICDM} 2010, Sydney, Australia, 14-17 December
  2010}}. \bibinfo{publisher}{{IEEE} Computer Society},
  \bibinfo{pages}{995--1000}.
\newblock


\bibitem[\protect\citeauthoryear{Rendle, Freudenthaler, Gantner, and
  Schmidt{-}Thieme}{Rendle et~al\mbox{.}}{2009}]%
        {DBLP:conf/uai/RendleFGS09}
\bibfield{author}{\bibinfo{person}{Steffen Rendle}, \bibinfo{person}{Christoph
  Freudenthaler}, \bibinfo{person}{Zeno Gantner}, {and} \bibinfo{person}{Lars
  Schmidt{-}Thieme}.} \bibinfo{year}{2009}\natexlab{}.
\newblock \showarticletitle{{BPR:} Bayesian Personalized Ranking from Implicit
  Feedback}. In \bibinfo{booktitle}{\emph{{UAI} 2009, Montreal, QC, Canada,
  June 18-21, 2009}}. \bibinfo{pages}{452--461}.
\newblock


\bibitem[\protect\citeauthoryear{Russakovsky, Deng, Su, Krause, Satheesh, Ma,
  Huang, Karpathy, Khosla, Bernstein, Berg, and Li}{Russakovsky
  et~al\mbox{.}}{2015}]%
        {DBLP:journals/ijcv/RussakovskyDSKS15}
\bibfield{author}{\bibinfo{person}{Olga Russakovsky}, \bibinfo{person}{Jia
  Deng}, \bibinfo{person}{Hao Su}, \bibinfo{person}{Jonathan Krause},
  \bibinfo{person}{Sanjeev Satheesh}, \bibinfo{person}{Sean Ma},
  \bibinfo{person}{Zhiheng Huang}, \bibinfo{person}{Andrej Karpathy},
  \bibinfo{person}{Aditya Khosla}, \bibinfo{person}{Michael~S. Bernstein},
  \bibinfo{person}{Alexander~C. Berg}, {and} \bibinfo{person}{Fei{-}Fei Li}.}
  \bibinfo{year}{2015}\natexlab{}.
\newblock \showarticletitle{ImageNet Large Scale Visual Recognition Challenge}.
\newblock \bibinfo{journal}{\emph{Int. J. Comput. Vis.}} \bibinfo{volume}{115},
  \bibinfo{number}{3} (\bibinfo{year}{2015}), \bibinfo{pages}{211--252}.
\newblock


\bibitem[\protect\citeauthoryear{Sarafianos, Giannakopoulos, Nikou, and
  Kakadiaris}{Sarafianos et~al\mbox{.}}{2017}]%
        {CL_CV_task_level}
\bibfield{author}{\bibinfo{person}{Nikolaos Sarafianos},
  \bibinfo{person}{Theodore Giannakopoulos}, \bibinfo{person}{Christophoros
  Nikou}, {and} \bibinfo{person}{Ioannis~A. Kakadiaris}.}
  \bibinfo{year}{2017}\natexlab{}.
\newblock \showarticletitle{Curriculum Learning for Multi-task Classification
  of Visual Attributes}. In \bibinfo{booktitle}{\emph{{ICCV} Workshops 2017,
  Venice, Italy, October 22-29, 2017}}. \bibinfo{publisher}{{IEEE} Computer
  Society}, \bibinfo{pages}{2608--2615}.
\newblock


\bibitem[\protect\citeauthoryear{Shi, Hu, Zhao, and Yu}{Shi
  et~al\mbox{.}}{2019}]%
        {DBLP:journals/tkde/ShiHZY19}
\bibfield{author}{\bibinfo{person}{Chuan Shi}, \bibinfo{person}{Binbin Hu},
  \bibinfo{person}{Wayne~Xin Zhao}, {and} \bibinfo{person}{Philip~S. Yu}.}
  \bibinfo{year}{2019}\natexlab{}.
\newblock \showarticletitle{Heterogeneous Information Network Embedding for
  Recommendation}.
\newblock \bibinfo{journal}{\emph{{IEEE} Trans. Knowl. Data Eng.}}
  \bibinfo{volume}{31}, \bibinfo{number}{2} (\bibinfo{year}{2019}),
  \bibinfo{pages}{357--370}.
\newblock


\bibitem[\protect\citeauthoryear{Shi, Kong, Huang, Yu, and Wu}{Shi
  et~al\mbox{.}}{2014}]%
        {DBLP:journals/tkde/ShiKHYW14}
\bibfield{author}{\bibinfo{person}{Chuan Shi}, \bibinfo{person}{Xiangnan Kong},
  \bibinfo{person}{Yue Huang}, \bibinfo{person}{Philip~S. Yu}, {and}
  \bibinfo{person}{Bin Wu}.} \bibinfo{year}{2014}\natexlab{}.
\newblock \showarticletitle{HeteSim: {A} General Framework for Relevance
  Measure in Heterogeneous Networks}.
\newblock \bibinfo{journal}{\emph{{IEEE} Trans. Knowl. Data Eng.}}
  \bibinfo{volume}{26}, \bibinfo{number}{10} (\bibinfo{year}{2014}),
  \bibinfo{pages}{2479--2492}.
\newblock
\urldef\tempurl%
\url{https://doi.org/10.1109/TKDE.2013.2297920}
\showDOI{\tempurl}


\bibitem[\protect\citeauthoryear{Shi, Li, Zhang, Sun, and Yu}{Shi
  et~al\mbox{.}}{2017}]%
        {DBLP:journals/tkde/ShiLZSY17}
\bibfield{author}{\bibinfo{person}{Chuan Shi}, \bibinfo{person}{Yitong Li},
  \bibinfo{person}{Jiawei Zhang}, \bibinfo{person}{Yizhou Sun}, {and}
  \bibinfo{person}{Philip~S. Yu}.} \bibinfo{year}{2017}\natexlab{}.
\newblock \showarticletitle{A Survey of Heterogeneous Information Network
  Analysis}.
\newblock \bibinfo{journal}{\emph{{IEEE} Trans. Knowl. Data Eng.}}
  \bibinfo{volume}{29}, \bibinfo{number}{1} (\bibinfo{year}{2017}),
  \bibinfo{pages}{17--37}.
\newblock


\bibitem[\protect\citeauthoryear{Sun, Han, Yan, Yu, and Wu}{Sun
  et~al\mbox{.}}{2011}]%
        {DBLP:journals/pvldb/SunHYYW11}
\bibfield{author}{\bibinfo{person}{Yizhou Sun}, \bibinfo{person}{Jiawei Han},
  \bibinfo{person}{Xifeng Yan}, \bibinfo{person}{Philip~S. Yu}, {and}
  \bibinfo{person}{Tianyi Wu}.} \bibinfo{year}{2011}\natexlab{}.
\newblock \showarticletitle{PathSim: Meta Path-Based Top-K Similarity Search in
  Heterogeneous Information Networks}.
\newblock \bibinfo{journal}{\emph{Proc. {VLDB} Endow.}} \bibinfo{volume}{4},
  \bibinfo{number}{11} (\bibinfo{year}{2011}), \bibinfo{pages}{992--1003}.
\newblock


\bibitem[\protect\citeauthoryear{Vaswani, Shazeer, Parmar, Uszkoreit, Jones,
  Gomez, Kaiser, and Polosukhin}{Vaswani et~al\mbox{.}}{2017}]%
        {DBLP:conf/nips/VaswaniSPUJGKP17}
\bibfield{author}{\bibinfo{person}{A. Vaswani}, \bibinfo{person}{N. Shazeer},
  \bibinfo{person}{N. Parmar}, \bibinfo{person}{J. Uszkoreit},
  \bibinfo{person}{L. Jones}, \bibinfo{person}{A.~N. Gomez},
  \bibinfo{person}{L. Kaiser}, {and} \bibinfo{person}{I. Polosukhin}.}
  \bibinfo{year}{2017}\natexlab{}.
\newblock \showarticletitle{Attention is All you Need}. In
  \bibinfo{booktitle}{\emph{{NeurIPS} 2017}}. \bibinfo{pages}{5998--6008}.
\newblock


\bibitem[\protect\citeauthoryear{Velickovic, Fedus, Hamilton, Li{\`{o}},
  Bengio, and Hjelm}{Velickovic et~al\mbox{.}}{2019}]%
        {DBLP:conf/iclr/VelickovicFHLBH19}
\bibfield{author}{\bibinfo{person}{Petar Velickovic}, \bibinfo{person}{William
  Fedus}, \bibinfo{person}{William~L. Hamilton}, \bibinfo{person}{Pietro
  Li{\`{o}}}, \bibinfo{person}{Yoshua Bengio}, {and} \bibinfo{person}{R.~Devon
  Hjelm}.} \bibinfo{year}{2019}\natexlab{}.
\newblock \showarticletitle{Deep Graph Infomax}. In
  \bibinfo{booktitle}{\emph{7th International Conference on Learning
  Representations, {ICLR} 2019, New Orleans, LA, USA, May 6-9, 2019}}.
\newblock


\bibitem[\protect\citeauthoryear{Wang, Ji, Shi, Wang, Ye, Cui, and Yu}{Wang
  et~al\mbox{.}}{2019b}]%
        {DBLP:conf/www/WangJSWYCY19}
\bibfield{author}{\bibinfo{person}{Xiao Wang}, \bibinfo{person}{Houye Ji},
  \bibinfo{person}{Chuan Shi}, \bibinfo{person}{Bai Wang},
  \bibinfo{person}{Yanfang Ye}, \bibinfo{person}{Peng Cui}, {and}
  \bibinfo{person}{Philip~S. Yu}.} \bibinfo{year}{2019}\natexlab{b}.
\newblock \showarticletitle{Heterogeneous Graph Attention Network}. In
  \bibinfo{booktitle}{\emph{The World Wide Web Conference, {WWW} 2019, San
  Francisco, CA, USA, May 13-17, 2019}}. \bibinfo{pages}{2022--2032}.
\newblock


\bibitem[\protect\citeauthoryear{Wang, Duan, Liao, Wu, and Zhuang}{Wang
  et~al\mbox{.}}{2019a}]%
        {DBLP:conf/aaai/WangDLWZ19}
\bibfield{author}{\bibinfo{person}{Yueyang Wang}, \bibinfo{person}{Ziheng
  Duan}, \bibinfo{person}{Binbing Liao}, \bibinfo{person}{Fei Wu}, {and}
  \bibinfo{person}{Yueting Zhuang}.} \bibinfo{year}{2019}\natexlab{a}.
\newblock \showarticletitle{Heterogeneous Attributed Network Embedding with
  Graph Convolutional Networks}. In \bibinfo{booktitle}{\emph{{AAAI} 2019,
  Honolulu, Hawaii, USA, January 27 - February 1, 2019}}.
  \bibinfo{pages}{10061--10062}.
\newblock


\bibitem[\protect\citeauthoryear{Wang, Tang, Lei, Song, Wang, and Zhang}{Wang
  et~al\mbox{.}}{2020}]%
        {DBLP:conf/cikm/WangTLSWZ20}
\bibfield{author}{\bibinfo{person}{Yifan Wang}, \bibinfo{person}{Suyao Tang},
  \bibinfo{person}{Yuntong Lei}, \bibinfo{person}{Weiping Song},
  \bibinfo{person}{Sheng Wang}, {and} \bibinfo{person}{Ming Zhang}.}
  \bibinfo{year}{2020}\natexlab{}.
\newblock \showarticletitle{DisenHAN: Disentangled Heterogeneous Graph
  Attention Network for Recommendation}. In \bibinfo{booktitle}{\emph{{CIKM}
  '20: The 29th {ACM} International Conference on Information and Knowledge
  Management, Virtual Event, Ireland, October 19-23, 2020}},
  \bibfield{editor}{\bibinfo{person}{Mathieu d'Aquin}, \bibinfo{person}{Stefan
  Dietze}, \bibinfo{person}{Claudia Hauff}, \bibinfo{person}{Edward Curry},
  {and} \bibinfo{person}{Philippe Cudr{\'{e}}{-}Mauroux}} (Eds.).
  \bibinfo{publisher}{{ACM}}, \bibinfo{pages}{1605--1614}.
\newblock
\urldef\tempurl%
\url{https://doi.org/10.1145/3340531.3411996}
\showDOI{\tempurl}


\bibitem[\protect\citeauthoryear{Yang, Xiao, Zhang, Sun, and Han}{Yang
  et~al\mbox{.}}{2020}]%
        {HNE_survey_han}
\bibfield{author}{\bibinfo{person}{Carl Yang}, \bibinfo{person}{Yuxin Xiao},
  \bibinfo{person}{Yu Zhang}, \bibinfo{person}{Yizhou Sun}, {and}
  \bibinfo{person}{Jiawei Han}.} \bibinfo{year}{2020}\natexlab{}.
\newblock \showarticletitle{Heterogeneous Network Representation Learning:
  Survey, Benchmark, Evaluation, and Beyond}.
\newblock \bibinfo{journal}{\emph{CoRR}}  \bibinfo{volume}{abs/2004.00216}
  (\bibinfo{year}{2020}).
\newblock
\urldef\tempurl%
\url{https://arxiv.org/abs/2004.00216}
\showURL{%
\tempurl}


\bibitem[\protect\citeauthoryear{Yu, Ren, Sun, Gu, Sturt, Khandelwal, Norick,
  and Han}{Yu et~al\mbox{.}}{2014}]%
        {DBLP:conf/wsdm/YuRSGSKNH14}
\bibfield{author}{\bibinfo{person}{Xiao Yu}, \bibinfo{person}{Xiang Ren},
  \bibinfo{person}{Yizhou Sun}, \bibinfo{person}{Quanquan Gu},
  \bibinfo{person}{Bradley Sturt}, \bibinfo{person}{Urvashi Khandelwal},
  \bibinfo{person}{Brandon Norick}, {and} \bibinfo{person}{Jiawei Han}.}
  \bibinfo{year}{2014}\natexlab{}.
\newblock \showarticletitle{Personalized entity recommendation: a heterogeneous
  information network approach}. In \bibinfo{booktitle}{\emph{{WSDM} 2014, New
  York, NY, USA, February 24-28, 2014}}. \bibinfo{pages}{283--292}.
\newblock


\bibitem[\protect\citeauthoryear{Yu, Ren, Sun, Sturt, Khandelwal, Gu, Norick,
  and Han}{Yu et~al\mbox{.}}{2013}]%
        {DBLP:conf/recsys/YuRSSKGNH13}
\bibfield{author}{\bibinfo{person}{Xiao Yu}, \bibinfo{person}{Xiang Ren},
  \bibinfo{person}{Yizhou Sun}, \bibinfo{person}{Bradley Sturt},
  \bibinfo{person}{Urvashi Khandelwal}, \bibinfo{person}{Quanquan Gu},
  \bibinfo{person}{Brandon Norick}, {and} \bibinfo{person}{Jiawei Han}.}
  \bibinfo{year}{2013}\natexlab{}.
\newblock \showarticletitle{Recommendation in heterogeneous information
  networks with implicit user feedback}. In \bibinfo{booktitle}{\emph{Seventh
  {ACM} Conference on Recommender Systems, RecSys '13, Hong Kong, China,
  October 12-16, 2013}}. \bibinfo{pages}{347--350}.
\newblock


\bibitem[\protect\citeauthoryear{Zhang, Song, Huang, Swami, and Chawla}{Zhang
  et~al\mbox{.}}{2019}]%
        {DBLP:conf/kdd/ZhangSHSC19}
\bibfield{author}{\bibinfo{person}{Chuxu Zhang}, \bibinfo{person}{Dongjin
  Song}, \bibinfo{person}{Chao Huang}, \bibinfo{person}{Ananthram Swami}, {and}
  \bibinfo{person}{Nitesh~V. Chawla}.} \bibinfo{year}{2019}\natexlab{}.
\newblock \showarticletitle{Heterogeneous Graph Neural Network}. In
  \bibinfo{booktitle}{\emph{{KDD} 2019, Anchorage, AK, USA, August 4-8, 2019}}.
  \bibinfo{pages}{793--803}.
\newblock


\bibitem[\protect\citeauthoryear{Zhang, Zhang, Xia, and Sun}{Zhang
  et~al\mbox{.}}{2020}]%
        {DBLP:journals/corr/abs-2001-05140}
\bibfield{author}{\bibinfo{person}{Jiawei Zhang}, \bibinfo{person}{Haopeng
  Zhang}, \bibinfo{person}{Congying Xia}, {and} \bibinfo{person}{Li Sun}.}
  \bibinfo{year}{2020}\natexlab{}.
\newblock \showarticletitle{Graph-Bert: Only Attention is Needed for Learning
  Graph Representations}.
\newblock \bibinfo{journal}{\emph{CoRR}}  \bibinfo{volume}{abs/2001.05140}
  (\bibinfo{year}{2020}).
\newblock
\showeprint[arxiv]{2001.05140}


\bibitem[\protect\citeauthoryear{Zhao, Zhou, Guan, Zhao, Ning, Qiu, and
  He}{Zhao et~al\mbox{.}}{2019}]%
        {DBLP:conf/kdd/0009ZGZNQH19}
\bibfield{author}{\bibinfo{person}{Jun Zhao}, \bibinfo{person}{Zhou Zhou},
  \bibinfo{person}{Ziyu Guan}, \bibinfo{person}{Wei Zhao}, \bibinfo{person}{Wei
  Ning}, \bibinfo{person}{Guang Qiu}, {and} \bibinfo{person}{Xiaofei He}.}
  \bibinfo{year}{2019}\natexlab{}.
\newblock \showarticletitle{IntentGC: {A} Scalable Graph Convolution Framework
  Fusing Heterogeneous Information for Recommendation}. In
  \bibinfo{booktitle}{\emph{{KDD} 2019, Anchorage, AK, USA, August 4-8, 2019}}.
  \bibinfo{pages}{2347--2357}.
\newblock


\bibitem[\protect\citeauthoryear{Zhou, Wang, Zhao, Zhu, Wang, Zhang, Wang, and
  Wen}{Zhou et~al\mbox{.}}{2020a}]%
        {DBLP:conf/cikm/ZhouWZZWZWW20}
\bibfield{author}{\bibinfo{person}{Kun Zhou}, \bibinfo{person}{Hui Wang},
  \bibinfo{person}{Wayne~Xin Zhao}, \bibinfo{person}{Yutao Zhu},
  \bibinfo{person}{Sirui Wang}, \bibinfo{person}{Fuzheng Zhang},
  \bibinfo{person}{Zhongyuan Wang}, {and} \bibinfo{person}{Ji{-}Rong Wen}.}
  \bibinfo{year}{2020}\natexlab{a}.
\newblock \showarticletitle{S3-Rec: Self-Supervised Learning for Sequential
  Recommendation with Mutual Information Maximization}. In
  \bibinfo{booktitle}{\emph{{CIKM} '20: The 29th {ACM} International Conference
  on Information and Knowledge Management, Virtual Event, Ireland, October
  19-23, 2020}}, \bibfield{editor}{\bibinfo{person}{Mathieu d'Aquin},
  \bibinfo{person}{Stefan Dietze}, \bibinfo{person}{Claudia Hauff},
  \bibinfo{person}{Edward Curry}, {and} \bibinfo{person}{Philippe
  Cudr{\'{e}}{-}Mauroux}} (Eds.). \bibinfo{publisher}{{ACM}},
  \bibinfo{pages}{1893--1902}.
\newblock
\urldef\tempurl%
\url{https://doi.org/10.1145/3340531.3411954}
\showDOI{\tempurl}


\bibitem[\protect\citeauthoryear{Zhou, Zhao, Bian, Zhou, Wen, and Yu}{Zhou
  et~al\mbox{.}}{2020b}]%
        {DBLP:conf/kdd/ZhouZBZWY20}
\bibfield{author}{\bibinfo{person}{Kun Zhou}, \bibinfo{person}{Wayne~Xin Zhao},
  \bibinfo{person}{Shuqing Bian}, \bibinfo{person}{Yuanhang Zhou},
  \bibinfo{person}{Ji{-}Rong Wen}, {and} \bibinfo{person}{Jingsong Yu}.}
  \bibinfo{year}{2020}\natexlab{b}.
\newblock \showarticletitle{Improving Conversational Recommender Systems via
  Knowledge Graph based Semantic Fusion}. In \bibinfo{booktitle}{\emph{{KDD}
  '20: The 26th {ACM} {SIGKDD} Conference on Knowledge Discovery and Data
  Mining, Virtual Event, CA, USA, August 23-27, 2020}},
  \bibfield{editor}{\bibinfo{person}{Rajesh Gupta}, \bibinfo{person}{Yan Liu},
  \bibinfo{person}{Jiliang Tang}, {and} \bibinfo{person}{B.~Aditya Prakash}}
  (Eds.). \bibinfo{publisher}{{ACM}}, \bibinfo{pages}{1006--1014}.
\newblock
\urldef\tempurl%
\url{https://doi.org/10.1145/3394486.3403143}
\showDOI{\tempurl}


\bibitem[\protect\citeauthoryear{Zhu, Xu, Yu, Liu, Wu, and Wang}{Zhu
  et~al\mbox{.}}{2020}]%
        {DBLP:journals/corr/abs-2006-04131}
\bibfield{author}{\bibinfo{person}{Yanqiao Zhu}, \bibinfo{person}{Yichen Xu},
  \bibinfo{person}{Feng Yu}, \bibinfo{person}{Qiang Liu}, \bibinfo{person}{Shu
  Wu}, {and} \bibinfo{person}{Liang Wang}.} \bibinfo{year}{2020}\natexlab{}.
\newblock \showarticletitle{Deep Graph Contrastive Representation Learning}.
\newblock \bibinfo{journal}{\emph{CoRR}}  \bibinfo{volume}{abs/2006.04131}
  (\bibinfo{year}{2020}).
\newblock


\end{thebibliography}

\ignore{
\appendix

\section{Research Methods}

\subsection{Part One}

Lorem ipsum dolor sit amet, consectetur adipiscing elit. Morbi
malesuada, quam in pulvinar varius, metus nunc fermentum urna, id
sollicitudin purus odio sit amet enim. Aliquam ullamcorper eu ipsum
vel mollis. Curabitur quis dictum nisl. Phasellus vel semper risus, et
lacinia dolor. Integer ultricies commodo sem nec semper.

\subsection{Part Two}

Etiam commodo feugiat nisl pulvinar pellentesque. Etiam auctor sodales
ligula, non varius nibh pulvinar semper. Suspendisse nec lectus non
ipsum convallis congue hendrerit vitae sapien. Donec at laoreet
eros. Vivamus non purus placerat, scelerisque diam eu, cursus
ante. Etiam aliquam tortor auctor efficitur mattis.

\section{Online Resources}

Nam id fermentum dui. Suspendisse sagittis tortor a nulla mollis, in
pulvinar ex pretium. Sed interdum orci quis metus euismod, et sagittis
enim maximus. Vestibulum gravida massa ut felis suscipit
congue. Quisque mattis elit a risus ultrices commodo venenatis eget
dui. Etiam sagittis eleifend elementum.

Nam interdum magna at lectus dignissim, ac dignissim lorem
rhoncus. Maecenas eu arcu ac neque placerat aliquam. Nunc pulvinar
massa et mattis lacinia.}

\end{document}